\documentclass[table,xcdraw,smallextended]{svjour3}

\usepackage{etex}
\usepackage{graphicx}
\usepackage{xcolor}
\usepackage[utf8]{inputenc}
\usepackage{listings}  
\usepackage{amsmath}
\usepackage{balance}
\usepackage{tabularx}
\usepackage{booktabs}
\usepackage{mdframed} 
\usepackage{verbatim} 
\usepackage{paralist}
\usepackage{listings}
\usepackage{courier}
\usepackage{xspace}
\usepackage{balance}
\usepackage{algorithm} 
\usepackage{algorithmic}
\usepackage{multirow}
\definecolor{ForestGreen}{RGB}{34,139,34}
\usepackage[]{pdfcomment}
\usepackage{lscape}

\usepackage{catoptions}

\usepackage{catoptions}
\makeatletter

\def\Autoref#1{%
  \begingroup
  \edef\reserved@a{\cpttrimspaces{#1}}%
  \ifcsndefTF{r@#1}{%
    \xaftercsname{\expandafter\testreftype\@fourthoffive}
      {r@\reserved@a}.\\{#1}%
  }{%
    \ref{#1}%
  }%
  \endgroup
}
\def\testreftype#1.#2\\#3{%
  \ifcsndefTF{#1autorefname}{%
    \def\reserved@a##1##2\@nil{%
      \uppercase{\def\ref@name{##1}}%
      \csn@edef{#1autorefname}{\ref@name##2}%
      \autoref{#3}%
    }%
    \reserved@a#1\@nil
  }{%
    \autoref{#3}%
  }%
}
\makeatother

\usepackage[inline]{enumitem}
\usepackage{amsmath}

\usepackage{amssymb}
\usepackage{pifont}
\newcommand{\cmark}{{\color{ForestGreen}\ding{51}}\xspace}%

\newcommand{\etal}{\textit{et al.}\xspace}
\newcommand{\ie}{\textit{i.e.}\xspace}
\newcommand{\eg}{\textit{e.g.}\xspace}
\newcommand{\gh}{GitHub\xspace}

\newcommand{\aampl}{\textsc{AAMPL}\xspace}

\newcommand{\sbampl}{\textsc{SBAMPL}\xspace}

\definecolor{ForestGreen}{RGB}{34,139,34}
\definecolor{diffstart}{RGB}{175,175,175}
\definecolor{diffincl}{RGB}{34,139,34}
\definecolor{diffrem}{RGB}{200,75,0}

\usepackage{listings}
\newcommand{\lstbg}[3][0pt]{{\fboxsep#1\colorbox{#2}{\strut #3}}}
\lstdefinelanguage{diff}{
  basicstyle=\scriptsize,
  morecomment=[f][\lstbg{red!20}]-,
  morecomment=[f][\lstbg{green!20}]+,
  morecomment=[f][\textit]{@@},
  breakatwhitespace=false,         
  breaklines=true,                 
  showtabs=false,                  
  tabsize=1                       
}

\newcommand{\TODO}[1]{\textcolor{red}{#1}\GenericError{}{LaTeX Error: TODO: #1}}\newcommand\todo\TODO
\newcommand{\rev}[1]{\textcolor{black}{#1}}
\newcommand{\revv}[1]{\textcolor{black}{#1}}

\newcommand{\RQ}[2]{RQ{#1}: #2}

\newcommand*\rotvertical{\rotatebox{90}}

\definecolor{red}{rgb}{1,0,0}
\definecolor{green}{rgb}{0,1,0}
\definecolor{blue}{rgb}{0,0,1}
\definecolor{cyan}{rgb}{0.4,1,1}
\definecolor{orange}{rgb}{0.9,0.5,0}
\definecolor{dkgreen}{rgb}{0,0.6,0}
\definecolor{gray}{rgb}{0.5,0.5,0.5}
\definecolor{purple}{rgb}{0.58,0,0.82}
\usepackage{times}

\newcommand{\done}[1]{{\color{blue}\bf\em DONE: #1}}\newcommand\DONE\done

\newcommand{\DCI}{DCI\xspace}
\newcommand{\DCIA}{DCI$_{AAMPL}$\xspace}
\newcommand{\DCII}{DCI$_{SBAMPL}$\xspace}

\usepackage[htt]{hyphenat}

\begin{document}

\title{An Approach and Benchmark to Detect Behavioral Changes of Commits in Continuous Integration}

\author{Benjamin Danglot, Martin Monperrus, Walter Rudametkin, Benoit Baudry}

\institute{B. Danglot \at
              Inria Lille - Nord Europe \\
                Lille - France\\
                \email{danglot@inria.fr}
           \and
           M. Monperrus \at
           KTH Royal Institute of Technology in Stockholm\\
           Stockholm - Sweden\\
           \email{martin.monperrus@csc.kth.se}
           \and
           W. Rudametkin\at
           University of Lille and INRIA\\
           Lille - France\\
           \email{oscar.vera-perez@inria.fr}
           \and
           B. Baudry \at
           KTH Royal Institute of Technology in Stockholm\\
           Stockholm - Sweden\\
           \email{baudry@kth.se}
}

\maketitle


\begin{abstract}
When a developer pushes a change to an application's codebase,  a good practice is to have a test case specifying this behavioral change. 
Thanks to continuous integration (CI), the test is run on subsequent commits to check that they do no introduce a regression for that behavior. 

In this paper, we propose an approach that detects behavioral changes in commits. As input, it takes a program, its test suite, and a commit. Its output is a set of test methods that capture the behavioral difference between the pre-commit and post-commit versions of the program. We call our approach DCI (Detecting behavioral changes in CI). It works by generating variations of the existing test cases through (i) assertion amplification and (ii) a search-based exploration of the input space. 

We evaluate our approach on a curated set of 60 commits from 6 open source Java projects.
To our knowledge, this is the first ever curated dataset of real-world behavioral changes.  Our evaluation shows that DCI is able to generate test methods that detect behavioral changes.
\revv{Our approach is fully automated and can be integrated into current development processes. The main limitations are that it targets unit tests and works on a relatively small fraction of commits. More specifically, DCI works on commits that have a unit test that already executes the modified code. In practice, from our benchmark projects, we found 15.29\% of commits to meet the conditions required by DCI.}
\end{abstract}


\section{Introduction}


In collaborative software projects, developers work in parallel on the same code base. 
Every time a developer integrates her changes, she submits them in the form of a \emph{commit} to a version control system.
The \emph{Continuous Integration} (CI) server~\cite{fowler2006continuous} merges the commit with the master branch, compiles and automatically runs the test suite to check that the commit behaves as expected.
Its ability to detect bugs early makes CI an essential contribution to quality assurance~\cite{Hilton:2016:UsageCI,duvall2007continuous}.

However, the effectiveness of Continuous Integration depends on one key property: each commit should include at least one test case $t_{new}$ that specifies the intended change.
For instance, assume one wants to integrate a bug fix.
In this case, the developer is expected to include a new test method, $t_{new}$, that specifies the program's desired behavior after the bug fix is applied.
This can be mechanically verified: $t_{new}$ should fail on the version of the code that does not include the fix (the \emph{pre-commit} version), and pass on the version that includes the fix (the \emph{post-commit} version).
However, many commits either do not include a $t_{new}$ or $t_{new}$ does not meet this fail/pass criterion.
The reason is that developers sometimes cut corners because of lack of time, expertise or discipline.
This is the problem we address in this paper.

In this paper, we aim to automatically generate test methods for each commit that is submitted to the CI.
In particular, we generate a test case $t_{gen}$ that specifies the behavioral change of each commit.
We consider a generated test case $t_{gen}$ to be relevant if it satisfies the following property: $t_{gen}$ \textit{passes} on the pre-commit version and \textit{fails} on the post-commit version.
To do so, we developed a new approach, called \DCI, that works in two steps.
First, we analyze the test cases of the pre-commit version and select the ones that exercise the parts of the code modified by the commit.
Second, our test generation techniques produce variant test cases that either add assertions~\cite{TaoXie2006} to existing tests or explore new inputs following a search-based test input generation approach~\cite{tonella}.
This process of automatic generation of $t_{gen}$ from existing tests is called \emph{test amplification}~\cite{zhang2012}.
We evaluate our approach on a benchmark of 60 commits selected from 6 open source Java projects, constructed with a novel and systematic methodology.
We analyzed 1576 commits and selected those that introduce a behavioral change (\eg, we do not want to generate tests for commits that only change comments).
We also make sure that all selected commits contain a developer-written test case that detects the behavioral change.
In our protocol, the developer's test case acts as a ground-truth to analyze the tests generated by \DCI.
Overall, we found 60 commits that satisfy the two essential properties we are looking for:
1) the commit introduces a behavioral change;
2) the commit has a human written test we can use for ground truth.
\revv{This corresponds to 15.3\% of commits in average.}
\rev{While this may appear to be a low proportion of commits, our approach is fully automated and developers can still benefit from its output without any manual intervention.}

To sum up, our contributions are:
\begin{itemize}
    \item \DCI (\textbf{D}etecting behavioral changes in \textbf{CI}), an approach based on \emph{test amplification} to generate new tests that detect the behavioral change introduced by a commit.
    \item An open-source implementation of \DCI for Java.
    \item A curated benchmark of 60 commits that introduce a behavioral change and include a test case to detect it, selected from 6 notable open source Java projects\footnote{\url{https://github.com/STAMP-project/dspot-experiments}}.
    \item A comprehensive evaluation based on 4 research questions that combines quantitative and qualitative analysis with manual assessment.
\end{itemize}

In \Autoref{sec:background} we motivate the need to have commits include a test case that specifies the behavioral change. 
In \Autoref{sec:techniques} we introduce our technical contribution: an approach for commit-based test selection and amplification. 
\Autoref{sec:evaluation} introduces our benchmark of commits, the evaluation protocol and the results of our experiments on 60 real commits.
\Autoref{sec:limitation} \rev{discusses the exact applicability scope of our approach.}
\Autoref{sec:threats} presents the threats validity and actions that have been taken to overcome them. 
In \Autoref{sec:related_work}, we expose the related work, their evaluation and the differences with our work and eventually we conclude in \Autoref{sec:conclusion}.

\section{Motivation \& Background}
\label{sec:background}

In this section, we introduce an example to motivate the need to generate new tests that specifically target the behavioral change introduced by a commit.
Then we introduce the key concepts on which we elaborate our solution to address this challenging test generation task.

\subsection{Motivating Example}

On August 10, a developer pushed a commit to the master branch of the XWiki-commons project. 
The change\footnote{\url{https://github.com/xwiki/xwiki-commons/commit/7e79f77}}, displayed in \Autoref{fig:motivating_example}, adds a comparison to ensure the equality of the objects returned by \texttt{getVersion()}.
The developer did not write a test method nor modify an existing one. 


\begin{lstlisting}[language=diff,caption=Commit \textsc{7e79f77} on XWiki-Commons that changes the behavior without a test.,label=fig:motivating_example]
@@ -260,7 +260,8 @@ public boolean equals(Object object)
} else {
    if (object instanceof FilterStreamType) {
        result = Objects.equals(getType(), ((FilterStreamType) object).getType())
-       && Objects.equals(getDataFormat(),
-                       ((FilterStreamType) object).getDataFormat());
+       && Objects.equals(getDataFormat(),
+                       ((FilterStreamType) object).getDataFormat())
+       && Objects.equals(getVersion(), 
+                       ((FilterStreamType) object).getVersion());
    } else {
        result = false;
    }
\end{lstlisting}

In this commit, the intent is to take into account the \texttt{version} (from method \texttt{getVersion}) in the \texttt{equals} method.
This change impacts the behavior of all methods that use it, \texttt{equals} being a highly used method.
Such a central behavioral change may impact the whole program, and the lack of a test case for this new behavior may have dramatic consequences in the future.
Without a test case, this change could be reverted and go undetected by the test suite and the Continuous Integration server, \ie the build would still pass.
Yet, a user of this program would encounter new errors, because of the changed behavior.
The developer took a risk when committing this change without a test case.

Our work on automatic test amplification in continuous integration aims at mitigating such risk: test amplification aims at ensuring that every commit include a new test method or a modification of an existing test method.
In this paper, we study how to automatically obtain a test method that highlights the behavioral change introduced by a commit.
This test method allows to identify the behavioral difference between the two versions of the program. 
Our goal is to use this new test method to ensure that any changed behavior can be caught in the future.

What we propose is as follows: 
when Continuous Integration is triggered, rather than just executing the test suite to find regressions, it could also run an analysis of the commit to know if it contains a behavioral change, in the form of a new method or the modification of an existing one.
If there is no appropriate test case to detect a behavioral change, our approach would provide one.
DCI would take as input the commit and a test suite, and generate a new test case that detects the behavioral change.

\subsection{Practibility}
\label{subsec:practicability}

We describe a complete scenario to sum up the vision of our approach's usage.

A developer commits a change into the program.
The Continuous Integration service is triggered;
the CI analyzes the commit.
There are two potential outcomes:

1) the developer provided a new test case or a modification to an existing one. 
In this case, the CI runs as usual, \eg it executes the test suite;

2) the developer did not provide a new test nor the modification of an existing one, the CI runs DCI on the commit to obtain a test method that detects the behavioral change and present it to the developer.

The developer can then validate the new test method that detects the behavioral change.
Following our definition, the new test method passes on the pre-commit version but fails on the post-commit version.
The current amplified test method cannot be added to the test suite, since it fails.
However, this test method is still useful, since one has only to negate the failing assertions, \eg change an \texttt{assertTrue} into an \texttt{assertFalse}, to obtain a valid and passing test method that explicitly executes the new behavior.
This can be done manually or automatically with approaches such as ReAssert\cite{ReAssert}.

\rev{\DCI could apply to any kind of test: unit-level or system-level.}
\rev{However, }from our experience, unit tests (vs integration tests) are the best target for \DCI, for two reasons.
First, they have a small scope, which allows \DCI to intensify its search, while an integration test, that contains a lot of code, would make \DCI explore the neighborhood in different ways.
Second, that is a consequence of the first, the unit tests are fast to execute compared to integration tests.
Since \DCI needs to execute the tests 5 times under amplification, it means that \DCI would be executed faster when it amplifies unit tests than when it amplified integration tests.

\DCI has been designed to be easy to use.
The only cost of \DCI is the time to set it up: in the ideal, happy-path case, it is meant to be a single command line through Maven goals.
Once \DCI is set up in continuous integration, it automatically runs at each commit and developers directly benefit from amplified test methods that strengthen the existing test suite.

%
%
\subsection{Behavioral Change}
\label{subsec:behavioral:change}

A \emph{behavioral change} is a source-code modification that triggers a new state for some inputs \cite{saff2004experimental}.
Considering the pre-commit version $P$ and the post-commit version $P'$ of a program, the commit introduces a behavioral change if it is possible to implement a test case that can trigger and observe the change, \ie, it passes on $P$ and fails on $P'$, or the opposite.
In short, the behavioral change must have an impact on the observable behavior of the program.

%
%
\subsection{Behavioral Change Detection}
\label{subsec:behavioral:change:detection}

Behavioral change detection is the task of identifying or generating a test or an input that distinguishes a behavioral change between two versions of the same program.
In this paper, we propose a novel approach to detect behavioral changes based on test amplification.

%
%
\subsection{Test Amplification}

Test amplification is the idea of improving existing tests with respect to a specific test criterion~\cite{zhang2012}.
We start from an existing test suite and create variant tests that improve a given test objective.
For instance, a test amplification tool may improve the code coverage of the test suite.
In this paper, our test objective is to improve the test suite's detection of behavioral changes introduced by commits.


\section{Behavioral Change Detection Approach}
\label{sec:techniques}

We propose an approach to produce test methods that detect the behavioral changes introduced by commits.
We call our approach \DCI (\textbf{D}etecting behavioral changes in \textbf{CI}), and propose to use it during continuous integration.

\subsection{Overview of DCI}
\label{subsec:global_overview}

\DCI takes as input a program, its test suite, and a commit modifying the program.
The commit, as done in version control systems, is basically the diff between two consecutive versions of the program.

\DCI outputs new test methods that detect the behavioral difference between the pre- and post-commit versions of the program.
The new tests pass on a given version, but fail on the other, demonstrating the presence of a behavioral change captured.

\begin{figure}
    \fbox{\includegraphics[width=.95\linewidth]{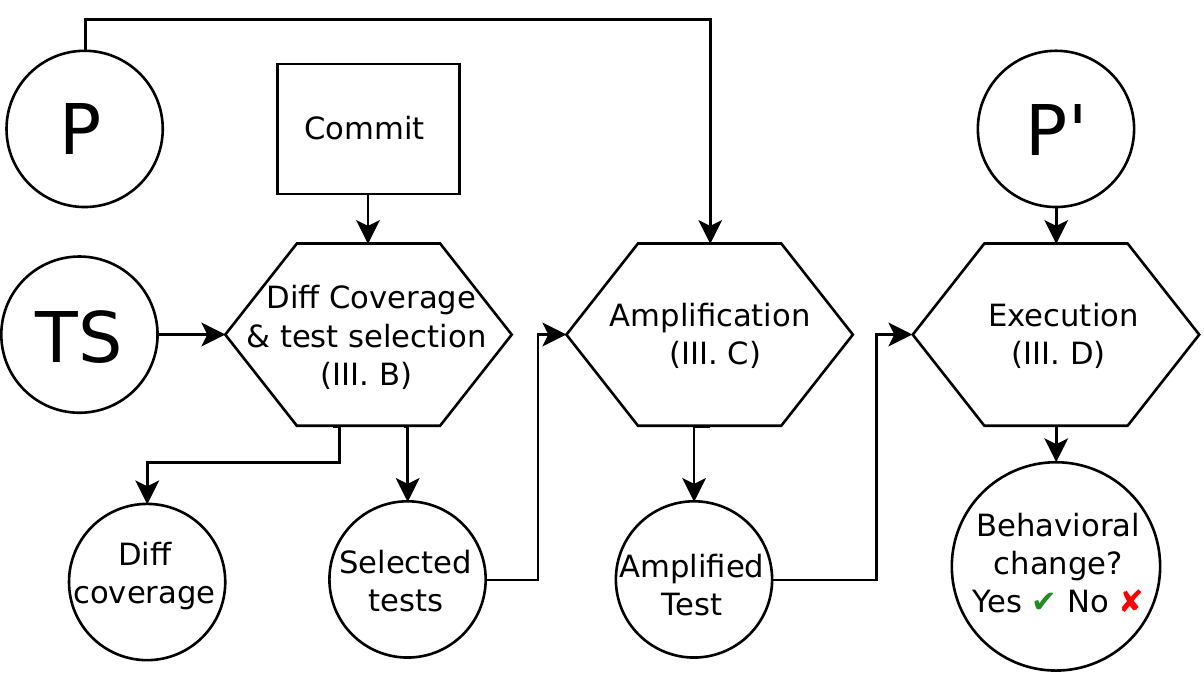}}
    \caption{Overview of our approach to detect behavioral changes in commits.}
    \label{fig:global_approach}
\end{figure}

\DCI computes the code coverage of the diff and selects test methods accordingly.
Then, it applies two kinds of test amplification to generate new test methods that detect the behavioral change.
\Autoref{fig:global_approach} sums up the different phases of the approach:

1) Compute the diff coverage and select the test methods to be amplified;

2) Amplify the selected tests based on the pre-commit version;

3) Execute amplified test methods against the post-commit version, and keep the failing test methods.

This process produces test methods that pass on the pre-commit version, fail on the post-commit version, hence they detect at least one behavioral change introduced by a given commit.

\subsection{Test Selection and Diff Coverage}
\label{subsec:compute_diff_coverage}

\DCI implements a feature that:
\begin{enumerate*}
\item reports the diff coverage of a commit, and
\item selects the set of unit tests that execute the diff.
\end{enumerate*}

To do so, \DCI first computes the code coverage for the whole test suite.

Second, it identifies the test methods that hit the statements modified by the diff. 

Third, it produces the two outcomes elicited earlier: 
the diff coverage, computed as the ratio of statements in the diff covered by the test suite over the total number of statements in the diff and 
the list of test methods that cover the diff.

Then, we select only test methods that are present in pre-commit version (\ie, we ignore the test methods added in the commit, if any).
The final list of test methods that cover the diff is then used to seed the amplification process.

\subsection{Test Amplification}

Once we have the initial tests that cover the diff, we want to make them detect the behavioral change and assess the new behavior.
This process of extending the scope of a test case is called test amplification~\cite{zhang2012}.
In \DCI, we build upon Xie's technique~\cite{TaoXie2006} and Tonella's evolutionary algorithm~\cite{tonella} to perform test amplification.

\subsubsection{Assertion Amplification}
\label{subsec:aampl}

A test method consists of a setup and assertions.
The former is responsible for putting the program under test into a specific state; the latter is responsible for verifying that the actual state of the program at the end of the test is the expected one.
To do this, assertions compare actual values against expected values: if the assertion holds, the program is considered correct, if not, the test case has revealed the presence of a bug.

Assertion amplification (\aampl) has been proposed by \cite{TaoXie2006}.
It takes as input a program and its test suite, and it synthesizes new assertions on public methods that capture the program state.
The targeted public methods are those that take no parameter, return a result, and match a Java naming convention of getters, \eg the method starts with \emph{get} or \emph{is}. The standard method \emph{toString()} is also used.
If a method used returns a complex Java Object, \aampl recursively uses getters on this object to generate deeper assertions.

In case the test method sets the program into an incorrect state and an exception is thrown, \aampl generates a test for this exception by wrapping the test method body in a \emph{try/catch} block. 
It also inserts a \emph{fail} statement at the end of the body of the \emph{try}, \ie it means that if the exception is not thrown the test method fails.

\begin{algorithm}[h]
\begin{algorithmic}[1]
\REQUIRE{Program $P$}
\REQUIRE{Test Suite $TS$}
\ENSURE{An Amplified Test Suite $ATS$}
\STATE{$ATS \leftarrow \emptyset$}
\FOR{$Test$ in $TS$}
    \STATE{$NoAssertTest \leftarrow removeAssertions(Test)$}
    \STATE{$InstrTest \leftarrow instrument(NoAssertTest)$}
    \STATE{$execute(InstrTest)$}
    \STATE{$AmplTest \leftarrow NoAssertTest.clone()$}
    \FOR{$Observ$ in $InstrTest.observations()$}
        \STATE{$Assert \leftarrow generateAssertion(Observ)$}
        \STATE{$AmplTest \leftarrow AmplTest.add(Assert)$}
    \ENDFOR
    \STATE{$ATS.add(select(AmplTest))$}
    \STATE{$ATS.add(AmplTest)$}
\ENDFOR
\RETURN $ATS$
\end{algorithmic}
\caption{\aampl: Assertion amplification algorithm.}
\label{algo:aampl}
\end{algorithm}

We present \aampl's pseudo-code in Algorithm \Autoref{algo:aampl}. First, it initializes an empty set of tests $ATS$ (Line 1). 
For each $Test$ method in the test suite $TS$ (Line 2), it removes the existing assertions to obtain $NoAssertTest$ (Line 3). 
Then, it instruments $NoAssertTest$ with observation points (Line 4) that allow retrieving values from the program at runtime, which results in $InstrTest$. 
In order to collect the values, it executes $InstrTest$ (Line 5).
Eventually, for each observation $Observ$ of the set of observations from $InstrTest$ (Line 7 to 10), it generates an assertion (Line 8) and adds it to the amplified tests $AmplTest$ (Line 9).
At the end, it selects amplified test according to a specific test criterion using the method $select()$ (Line 11) and add selected amplified test methods to the set of test methods $AmplTest$, in other words, an amplified test suite (Line 13).

To sum up, \aampl increases the number of assertions. 
By construction, it specifies more behaviors than the original test suite.
DCI$_{AAMPL}$ is the \aampl mode for DCI.

%
%
\subsubsection{Search-based Amplification}
\label{subsec:sbampl}
Search-based test amplification consists in running stochastic transformations on test code~\cite{tonella}.
For \DCII, this process consists in

a) generating a set of original test methods by applying code transformations;

b) executing \aampl to synthesize new assertions for these test methods for which the input has been modified at the previous step;

c) repeating this process $nb$ times\footnote{by default, $nb=3$}, each time seeding with the previously amplified test methods.

This final step allows the search-based algorithm to explore more inputs, and thus improve the chance of triggering new behaviors.

\begin{algorithm}[h]
\begin{algorithmic}[1]
\REQUIRE{Program $P$}
\REQUIRE{Program $P'$}
\REQUIRE{Test Suite $TS$}
\REQUIRE{Iterations number $Nb$}
\ENSURE{An Amplified Test Suite $ATS$}
\STATE{$ATS \leftarrow \emptyset$}
\STATE{$TmpTests \leftarrow \emptyset$}
\FOR{$Test$ in $TS$}
    \STATE{$TmpTests \leftarrow Test$}
    \FOR{$i \leftarrow 0, i < Nb$}
        \STATE{$TransformedTests \leftarrow transform(TmpTests)$}
        \STATE{$AmplifiedTests \leftarrow aampl(TransformedTests)$}
        \STATE{$ATS.add(select(AmplifiedTests))$}
        \STATE{$TmpTests \leftarrow AmplifiedTests$}
    \ENDFOR
\ENDFOR
\RETURN $ATS$
\end{algorithmic}
\caption{\sbampl: Search based amplification algorithm}
\label{algo:sbampl}
\end{algorithm}

We present the search-based amplification algorithm in Algorithm \Autoref{algo:sbampl}.
This algorithm is a basic Hill Climbing algorithm.
It takes as input a program with two distinct versions $P$ and $P'$, its test suite $TS$ and a number of iterations $nb$, (in our case $nb=3$).
It produces an amplified test suite that contains test methods that pass on $P$ but fail on $P'$.
To do so, it initializes an empty set of amplified test methods $ATS$ (Line 1), which will be the final output, and $TmpTests$ (Line 2) which is a temporary set.
Then, for each test method in the test suite $TS$ (Line 3), it applies the following operations:

1) transform the current set of test methods (Line 6) to obtain $TransformedTests$;

2) apply \aampl on $TransformedTests$ (Line 7, see \Autoref{algo:aampl}) to obtain $AmplifiedTests$; 

3) select amplified test methods using the method $select()$, and add them to $ATS$ 
(the method $select()$ executes the amplified tests on $P'$ and keeps only tests that fail, \ie that detect a behavioral change);

and Finally, 4) affects $AmplifiedTests$ to $TmpTests$ in order to stack transformations.

\begin{table}
	\caption{Test transformations considered in our study}
	\centering
	\begin{tabular}{c|l}
		\toprule
		Types & Operators\\
		\midrule
		Number &
		\begin{tabular}{l}
			add 1 to an integer\\
			minus 1 to an integer\\
			replace an integer by zero\\
			replace an integer by the maximum value (Integer.MAX\_VALUE in Java)\\
			replace an integer by the minimum value (\texttt{Integer.MIN\_VALUE} in Java).\\
		\end{tabular}\\
		\midrule
		Boolean &  	
		\begin{tabular}{l}
			negate the value.
		\end{tabular}\\
		\midrule
		String &
		\begin{tabular}{l}
			replace a string with another existing string.\\
			replace a string with white space, or a system path separator, or a system file separator.\\
			add 1 random character to the string.\\
			remove 1 random character from the string.\\
			replace 1 random character in the string by another random character.\\
			replace the string with a random string of the same size.\\
			replace the string with the \texttt{null} value.\\
		\end{tabular}\\
		\bottomrule
	\end{tabular}
	\label{table:test-transformations}
\end{table}

In our study, we consider the test transformations in \Autoref{table:test-transformations}.


\DCII is the search-based amplification mode for DCI.

\subsection{Execution and Change Detection}
\label{sec:change-detection}

The final step performed by \DCI consists in checking whether that the amplified test methods detect behavioral changes.
Because \DCI amplifies test methods using the pre-commit version, all amplified test methods pass on this version, by construction. 
Consequently, for the last step, \DCI runs the amplified test methods only on the post-commit version. 
Every test that fails is in fact detecting a behavioral change introduced by the commit, and is a success. 
DCI keeps the tests that successfully detect behavioral changes.

\subsection{Implementation}
\label{sub:implementation}

\DCI is implemented in Java and is built on top of the OpenClover and Gumtree~\cite{falleri:hal-01054552} libraries.
It computes the global coverage of the test suite with OpenClover, which instruments and executes the test suite.
Then, it uses Gumtree to have an AST representation of the diff.
\DCI matches the diff with the test that executes those lines. 
Through its Maven plugin, \DCI can be seamlessly implemented into continuous integration.
\DCI is publicly available on \gh.\footnote{\url{https://github.com/STAMP-project/dspot.git}}

\section{Evaluation}
\label{sec:evaluation}

To evaluate the \DCI approach, we design an experimental protocol to answer the following research questions:

\newcommand{\rqdetection}{\RQ{1}{To what extent are DCI$_{AAMPL}$ and DCI$_{SBAMPL}$ able to produce amplified test methods that detect the behavioral changes?}}
\newcommand{\rqselection}{\RQ{3}{What is the effectiveness of our test selection method?}}
\newcommand{\rqhuman}{\RQ{4}{How do human and generated tests that detect behavioral changes differ?}}
\newcommand{\rqiteration}{\RQ{2}{What  is the impact of the number of iteration performed by  DCI$_{SBAMPL}$?}}

\noindent
\begin{itemize}
\item \rqdetection
\item \rqiteration
\item \rqselection
\item \rqhuman
\end{itemize}

\subsection{Benchmark}
\label{sec:benchmark}
To the best of our knowledge, there is no benchmark of commits in Java with real behavioral changes in the literature.
Consequently, we devise a project and commit selection procedure in order to construct a benchmark for our approach.

\paragraph{Project selection}
We need software projects that are

1) publicly-available,

2) written in Java,

3) and use continuous integration.

We pick the projects from the dataset in \cite{descartes} and \cite{dspot-emse}, which is composed of mature Java projects from \gh.

\paragraph{Commit selection}
We take commits in inverse chronological order, from newest to oldest.
On September 10 2018, we selected the first 10 commits that match the following criteria:

- The commit modifies Java files (most behavioral changes are source code changes.\footnote{We are aware that behavioral changes can be introduced in other ways, such as modifying dependencies or configuration files \cite{Test:Coverage:Evolution}.}).

- The changes of the commit must be covered by the pre-commit test suite.
To do so, we compute the diff coverage. 
If the coverage is 0\%, we discard the commit. 
We do this because if the change is not covered, we cannot select any test methods to be amplified, which is what we want to evaluate.

- The commit provides or modifies a manually written test that detects a behavioral change. 
To verify this property, we execute the test on the pre-commit version. 
If it fails, it means that the test detects at least 1 behavioral change.
We will use this test as a \textit{ground-truth test} in \textbf{RQ4}.

Together, these criteria ensure that all selected commits:

1) modify java files,

2) that there is at least 1 test in the pre-commit version of the program that executes the diff and can be used to seed the amplification process

3) provide or modify a manually written test case that detects a behavioral change (which will be used as ground-truth for comparing generated tests), and

4) There is no structural change in the commit between both versions, \eg no change in method signature and deletion of classes (this is ensured since the pre-commit test suite compiles and runs against the post-commit version of the program and vice-versa.)

\paragraph{Final benchmark}
\begin{table}[h]
\centering
\def\arraystretch{1}
\setlength\tabcolsep{4pt}
\caption{Considered Period for Selecting Commits.}
\begin{tabular}{lc|rr|cccc}
\hline
project &
LOC &
\begin{tabular}{c}start\\date\end{tabular}&
\begin{tabular}{c}end\\date\end{tabular}&
\begin{tabular}{c}\#total\\commits\end{tabular}&
\begin{tabular}{c}\#matching\\commits\end{tabular}&
\begin{tabular}{c}\#selected\\commits\end{tabular}\\
\hline
\scriptsize{commons-io}	&	59607	&	9/10/2015	&	9/29/2018	&	385	&	49 / 12.73\%	&	10	\\
\rowcolor[HTML]{EFEFEF}
\scriptsize{commons-lang}	&	77410	&	11/22/2017	&	10/9/2018	&	227	&	40 / 17.62\%	&	10	\\
\scriptsize{gson}	&	49766	&	6/14/2016	&	10/9/2018	&	159	&	56 / 35.22\%	&	10	\\
\rowcolor[HTML]{EFEFEF}
\scriptsize{jsoup}	&	20088	&	12/21/2017	&	10/10/2018	&	50	&	42 / 84.00\%    &	10	\\
\scriptsize{mustache.java}	&	10289	&	7/6/2016	&	04/18/2019	&	68	&	28 / 41.18\%	&	10	\\
\rowcolor[HTML]{EFEFEF}
\scriptsize{xwiki-commons}	&	87289	&	10/31/2017	&	9/29/2018	&	687	&	26 / 3.78\%	&	10	\\
\hline
\scriptsize{summary}	&	304449	&	9/10/2015	&	04/18/2019	&	avg(262.67)	&	avg(40.17 / 15.29\%)	&	60	\\
\end{tabular}
\label{tab:benchmark}
\end{table}

\Autoref{tab:benchmark} shows the main descriptive statistics of the benchmark dataset.
The \emph{project} column is the name of the project.
The \emph{LOC} column is the number of lines of code computed with \textit{cloc}.
The \emph{start date} column is the date of the project's oldest commit.
The \emph{end date} column is the date of the project's newest commit.
\revv{The \emph{\#total commit} column is the total number of commits we analyzed.}
%
\revv{\emph{\#Matching commits} is the number of commits that match our first two criteria to run DCI but might not provide a test in the post-commit version that fails on the pre-commit version of the program.}
\revv{We could potentially apply DCI on all \emph{\#matching commits}, but for this paper, we cannot validate DCI with them because they might not provide a ground-truth test.}
\revv{The \emph{\#selected commits} column shows the number of commits we select for evaluation.}
\revv{It is a subset of \emph{\#matching commits} from which we searched for the first 10 commits per project that match all criteria, including a ground-truth test to evaluate DCI.}
%
The bottom row reports a summary of the benchmark dataset with the total number of lines of code, 
the oldest and the newest commit dates, 
the average number of commits analyzed, 
the average number of commits matching all the criteria but the third:
there is a test in the post-commit version of the program that detect the behavioral change, 
and the total number of selected commits.
The percentage in parenthesis next to the averages are percentage of averages, \eg $\frac{\#matching}{\#total}$.
We note that our benchmark is only composed of recent commits from notable open-source projects and is available on \gh at \url{https://github.com/STAMP-project/dspot-experiments}.

\subsection{Protocol}
\label{subsec:protocol}

To answer \textbf{RQ1}, we run DCI$_{AAMPL}$ and DCI$_{SBAMPL}$ on the benchmark projects.
We then report the total number of behavioral changes successfully detected by DCI, \ie the number of commits for which DCI generates at least 1 test method that passes on the pre-commit version but fails on the post-commit version.
We also discuss 1 case study of a successful behavioral change detection.

To answer \textbf{RQ2}, 
we run DCI$_{SBAMPL}$ for 1, 2 and 3 iterations on the benchmark projects. We report the number of behavioral changes successfully detected for each number of iterations in the main loop.
\rev{In addition, we want to have a proper understanding of the impact of randomness as follows.
We consider the case of $n=1$ iteration. 
For "$n=1$, we run DCI for each commit for 10 different seeds in addition to the reference run with the default seed, totalling 11 runs.}.
\rev{From those runs, we compute the confidence interval on the number of successes, \ie the number of time \DCI generates at least one amplified test method that detects the behavioral change, in order to measure the uncertainty of the result.}
\rev{To do this, we use Python libraries \emph{scipy} and \emph{numpy}, and we consider a confidence level of 95\%. Per our open-science approach, the interested reader has access to both the raw data and the script computing the confidence interval. 
}\footnote{\url{https://github.com/STAMP-project/dspot-experiments/tree/master/src/main/python/april-2019}}

For \textbf{RQ3}, the test selection method is considered effective if the tests selected for amplification semantically relate to the code changed by the commit. 
To assess this, we perform a manual analysis.
We randomly select 1 commit per project in the benchmark, and we manually analyze whether the automatically selected tests for this commit are semantically related to the behavioral changes in the commit. 

To answer \textbf{RQ4}, we use the ground-truth tests written or modified by developers in the selected commits.
We manually compare the amplified test methods that detect behavioral changes to the human tests, for 1 commit per project.

\subsection{Results}
\label{subsec:result}
\begin{table}
\centering
\small
\def\arraystretch{0.6}
\setlength\tabcolsep{0.5pt} 
\caption{Performance evaluation of DCI on 60 commits from 6 large open-source projects.}
\label{tab:overall_result}
\begin{tabular}{l|crccccccccccccc}
&
id&
date&
\#Test&
\begin{tabular}{c}\#Modified\\Tests\end{tabular}&
{\color{ForestGreen}{+\xspace}} / {\color{red}{-\xspace}}&
Cov&
\begin{tabular}{c}\#Selected\\Tests\end{tabular}&
\begin{tabular}{c}\#\aampl\\Tests\end{tabular}&
Time&
\begin{tabular}{c}\#\sbampl\\Tests\end{tabular}&
Time\\\\
\midrule
\multirow{11}{*}{\rotvertical{commons-io}}
&  c6b8a38  &  6/12/18 &  1348  &  2  &  {\color{ForestGreen}{104\xspace}} / {\color{red}{3\xspace}}  &  100.0  &  3  &  0  &  10.0s  &  0  &  98.0s\\
&  2736b6f  &  12/21/17 &  1343  &  2  &  {\color{ForestGreen}{164\xspace}} / {\color{red}{1\xspace}}  &  1.79  &  8  &  0  &  19.0s  &  \cmark(12)  &  76.3m\\
&  a4705cc  &  4/29/18 &  1328  &  1  &  {\color{ForestGreen}{37\xspace}} / {\color{red}{0\xspace}}  &  100.0  &  2  &  0  &  10.0s  &  0  &  38.1m\\
&  f00d97a  &  5/2/17 &  1316  &  10  &  {\color{ForestGreen}{244\xspace}} / {\color{red}{25\xspace}}  &  100.0  &  2  &  \cmark(1)  &  10.0s  &  \cmark(39)  &  27.0s\\
&  3378280  &  4/25/17 &  1309  &  2  &  {\color{ForestGreen}{5\xspace}} / {\color{red}{5\xspace}}  &  100.0  &  1  &  \cmark(1)  &  9.0s  &  \cmark(11)  &  24.0s\\
&  703228a  &  12/2/16 &  1309  &  1  &  {\color{ForestGreen}{6\xspace}} / {\color{red}{0\xspace}}  &  50.0  &  8  &  0  &  19.0s  &  0  &  71.0m\\
&  a7bd568  &  9/24/16 &  1163  &  1  &  {\color{ForestGreen}{91\xspace}} / {\color{red}{83\xspace}}  &  50.0  &  8  &  0  &  20.0s  &  0  &  65.2m\\
&  81210eb  &  6/2/16 &  1160  &  1  &  {\color{ForestGreen}{10\xspace}} / {\color{red}{2\xspace}}  &  100.0  &  1  &  0  &  8.0s  &  \cmark(8)  &  23.0s\\
&  57f493a  &  11/19/15 &  1153  &  1  &  {\color{ForestGreen}{15\xspace}} / {\color{red}{1\xspace}}  &  100.0  &  8  &  0  &  7.0s  &  0  &  54.0s\\
&  5d072ef  &  9/10/15 &  1125  &  12  &  {\color{ForestGreen}{74\xspace}} / {\color{red}{34\xspace}}  &  68.42  &  25  &  \cmark(6)  &  29.0s  &  \cmark(1538)  &  2.2h\\
\midrule
\rowcolor[HTML]{EFEFEF}
&  total  &  \xspace{} &  \xspace{}  &  \xspace{}  &  \xspace{}  &  \xspace{}  &  66  &  8  &  2.4m  &  1608  &  6.5h\\
\midrule
&  average  &  \xspace{} &  \xspace{}  &  \xspace{}  &  \xspace{}  &  \xspace{}  &  6.60  &  0.80  &  14.5s  &  160.80  &  38.8m\\
\midrule
\multirow{11}{*}{\rotvertical{commons-lang}}
&  f56931c  &  7/2/18 &  4105  &  1  &  {\color{ForestGreen}{30\xspace}} / {\color{red}{4\xspace}}  &  25.0  &  42  &  0  &  2.4m  &  0  &  8.5m\\
&  87937b2  &  5/22/18 &  4101  &  1  &  {\color{ForestGreen}{114\xspace}} / {\color{red}{0\xspace}}  &  77.78  &  16  &  0  &  35.0s  &  0  &  18.1m\\
&  09ef69c  &  5/18/18 &  4100  &  1  &  {\color{ForestGreen}{10\xspace}} / {\color{red}{1\xspace}}  &  100.0  &  4  &  0  &  16.0s  &  0  &  98.8m\\
&  3fadfdd  &  5/10/18 &  4089  &  1  &  {\color{ForestGreen}{7\xspace}} / {\color{red}{1\xspace}}  &  100.0  &  9  &  0  &  17.0s  &  \cmark(4)  &  17.2m\\
&  e7d16c2  &  5/9/18 &  4088  &  1  &  {\color{ForestGreen}{13\xspace}} / {\color{red}{1\xspace}}  &  33.33  &  7  &  0  &  16.0s  &  \cmark(2)  &  15.1m\\
&  50ce8c4  &  3/8/18 &  4084  &  4  &  {\color{ForestGreen}{40\xspace}} / {\color{red}{1\xspace}}  &  90.91  &  2  &  \cmark(1)  &  28.0s  &  \cmark(135)  &  2.0m\\
&  2e9f3a8  &  2/11/18 &  4084  &  2  &  {\color{ForestGreen}{79\xspace}} / {\color{red}{4\xspace}}  &  30.0  &  47  &  0  &  79.0s  &  0  &  66.5m\\
&  c8e61af  &  2/10/18 &  4082  &  1  &  {\color{ForestGreen}{8\xspace}} / {\color{red}{1\xspace}}  &  100.0  &  10  &  0  &  17.0s  &  0  &  16.0s\\
&  d8ec011  &  11/12/17 &  4074  &  1  &  {\color{ForestGreen}{11\xspace}} / {\color{red}{1\xspace}}  &  100.0  &  5  &  0  &  31.0s  &  0  &  2.3m\\
&  7d061e3  &  11/22/17 &  4073  &  1  &  {\color{ForestGreen}{16\xspace}} / {\color{red}{1\xspace}}  &  100.0  &  8  &  0  &  17.0s  &  0  &  11.4m\\
\midrule
\rowcolor[HTML]{EFEFEF}
&  total  &  \xspace{} &  \xspace{}  &  \xspace{}  &  \xspace{}  &  \xspace{}  &  150  &  1  &  6.7m  &  141  &  4.0h\\
\midrule
&  average  &  \xspace{} &  \xspace{}  &  \xspace{}  &  \xspace{}  &  \xspace{}  &  15.00  &  0.10  &  40.5s  &  14.10  &  24.0m\\
\midrule
\multirow{11}{*}{\rotvertical{gson}}
&  b1fb9ca  &  9/22/17 &  1035  &  1  &  {\color{ForestGreen}{23\xspace}} / {\color{red}{0\xspace}}  &  50.0  &  166  &  0  &  4.2m  &  0  &  92.5m\\
&  7a9fd59  &  9/18/17 &  1033  &  2  &  {\color{ForestGreen}{21\xspace}} / {\color{red}{2\xspace}}  &  83.33  &  14  &  0  &  15.0s  &  \cmark(108)  &  2.1m\\
&  03a72e7  &  8/1/17 &  1031  &  2  &  {\color{ForestGreen}{43\xspace}} / {\color{red}{11\xspace}}  &  68.75  &  371  &  0  &  7.7m  &  0  &  3.2h\\
&  74e3711  &  6/20/17 &  1029  &  1  &  {\color{ForestGreen}{68\xspace}} / {\color{red}{5\xspace}}  &  8.0  &  1  &  0  &  4.0s  &  0  &  16.0s\\
&  ada597e  &  5/31/17 &  1029  &  2  &  {\color{ForestGreen}{28\xspace}} / {\color{red}{3\xspace}}  &  100.0  &  5  &  0  &  8.0s  &  0  &  8.7m\\
&  a300148  &  5/31/17 &  1027  &  7  &  {\color{ForestGreen}{103\xspace}} / {\color{red}{2\xspace}}  &  18.18  &  665  &  0  &  9.2m  &  \cmark(6)  &  4.9h\\
&  9a24219  &  4/19/17 &  1019  &  1  &  {\color{ForestGreen}{13\xspace}} / {\color{red}{1\xspace}}  &  100.0  &  36  &  0  &  2.2m  &  0  &  48.9m\\
&  9e6f2ba  &  2/16/17 &  1018  &  2  &  {\color{ForestGreen}{56\xspace}} / {\color{red}{2\xspace}}  &  50.0  &  9  &  0  &  32.0s  &  \cmark(2)  &  8.5m\\
&  44cad04  &  11/26/16 &  1015  &  1  &  {\color{ForestGreen}{6\xspace}} / {\color{red}{0\xspace}}  &  100.0  &  2  &  0  &  15.0s  &  \cmark(37)  &  40.0s\\
&  b2c00a3  &  6/14/16 &  1012  &  4  &  {\color{ForestGreen}{242\xspace}} / {\color{red}{29\xspace}}  &  60.71  &  383  &  0  &  7.9m  &  0  &  3.6h\\
\midrule
\rowcolor[HTML]{EFEFEF}
&  total  &  \xspace{} &  \xspace{}  &  \xspace{}  &  \xspace{}  &  \xspace{}  &  1652  &  0  &  32.4m  &  153  &  14.4h\\
\midrule
&  average  &  \xspace{} &  \xspace{}  &  \xspace{}  &  \xspace{}  &  \xspace{}  &  165.20  &  0.00  &  3.2m  &  15.30  &  86.5m\\
\midrule
\multirow{11}{*}{\rotvertical{jsoup}}
&  426ffe7  &  5/11/18 &  668  &  4  &  {\color{ForestGreen}{27\xspace}} / {\color{red}{46\xspace}}  &  64.71  &  27  &  \cmark(2)  &  42.0s  &  \cmark(198)  &  33.6m\\
&  a810d2e  &  4/29/18 &  666  &  1  &  {\color{ForestGreen}{27\xspace}} / {\color{red}{1\xspace}}  &  80.0  &  5  &  0  &  10.0s  &  0  &  26.6m\\
&  6be19a6  &  4/29/18 &  664  &  1  &  {\color{ForestGreen}{23\xspace}} / {\color{red}{1\xspace}}  &  50.0  &  50  &  0  &  69.0s  &  0  &  67.7m\\
&  e38dfd4  &  4/28/18 &  659  &  1  &  {\color{ForestGreen}{66\xspace}} / {\color{red}{15\xspace}}  &  90.0  &  18  &  0  &  35.0s  &  0  &  12.5m\\
&  e9feec9  &  4/15/18 &  654  &  1  &  {\color{ForestGreen}{15\xspace}} / {\color{red}{3\xspace}}  &  100.0  &  4  &  0  &  9.0s  &  0  &  95.0s\\
&  0f7e0cc  &  4/14/18 &  653  &  2  &  {\color{ForestGreen}{56\xspace}} / {\color{red}{15\xspace}}  &  84.62  &  330  &  0  &  6.5m  &  \cmark(36)  &  11.8h\\
&  2c4e79b  &  4/14/18 &  650  &  2  &  {\color{ForestGreen}{82\xspace}} / {\color{red}{2\xspace}}  &  50.0  &  44  &  0  &  67.0s  &  0  &  4.7h\\
&  e5210d1  &  12/22/17 &  647  &  1  &  {\color{ForestGreen}{3\xspace}} / {\color{red}{3\xspace}}  &  100.0  &  14  &  0  &  9.0s  &  0  &  4.9m\\
&  df272b7  &  12/22/17 &  647  &  2  &  {\color{ForestGreen}{17\xspace}} / {\color{red}{1\xspace}}  &  100.0  &  13  &  0  &  9.0s  &  0  &  4.6m\\
&  3676b13  &  12/21/17 &  648  &  6  &  {\color{ForestGreen}{104\xspace}} / {\color{red}{12\xspace}}  &  38.46  &  239  &  0  &  6.2m  &  \cmark(52)  &  6.8h\\
\midrule
\rowcolor[HTML]{EFEFEF}
&  total  &  \xspace{} &  \xspace{}  &  \xspace{}  &  \xspace{}  &  \xspace{}  &  744  &  2  &  16.8m  &  286  &  25.8h\\
\midrule
&  average  &  \xspace{} &  \xspace{}  &  \xspace{}  &  \xspace{}  &  \xspace{}  &  74.40  &  0.20  &  101.0s  &  28.60  &  2.6h\\
\midrule
\multirow{11}{*}{\rotvertical{mustache.java}}
&  a1197f7  &  1/25/18 &  228  &  1  &  {\color{ForestGreen}{43\xspace}} / {\color{red}{57\xspace}}  &  77.78  &  131  &  0  &  11.8m  &  \cmark(204)  &  10.1h\\
&  8877027  &  11/19/17 &  227  &  1  &  {\color{ForestGreen}{22\xspace}} / {\color{red}{2\xspace}}  &  33.33  &  47  &  0  &  7.3m  &  0  &  100.2m\\
&  d8936b4  &  2/1/17 &  219  &  2  &  {\color{ForestGreen}{46\xspace}} / {\color{red}{6\xspace}}  &  60.0  &  168  &  0  &  12.7m  &  0  &  84.2m\\
&  88718bc  &  1/25/17 &  216  &  2  &  {\color{ForestGreen}{29\xspace}} / {\color{red}{1\xspace}}  &  100.0  &  1  &  \cmark(1)  &  7.0s  &  \cmark(149)  &  3.7m\\
&  339161f  &  9/23/16 &  214  &  2  &  {\color{ForestGreen}{32\xspace}} / {\color{red}{10\xspace}}  &  77.78  &  123  &  0  &  8.6m  &  \cmark(1312)  &  5.8h\\
&  774ae7a  &  8/10/16 &  214  &  2  &  {\color{ForestGreen}{17\xspace}} / {\color{red}{2\xspace}}  &  100.0  &  11  &  0  &  66.0s  &  \cmark(124)  &  6.8m\\
&  94847cc  &  7/29/16 &  214  &  2  &  {\color{ForestGreen}{17\xspace}} / {\color{red}{2\xspace}}  &  100.0  &  95  &  0  &  11.5m  &  \cmark(2509)  &  21.4h\\
&  eca08ca  &  7/14/16 &  212  &  4  &  {\color{ForestGreen}{47\xspace}} / {\color{red}{10\xspace}}  &  80.0  &  18  &  0  &  87.0s  &  0  &  41.8m\\
&  6d7225c  &  7/7/16 &  212  &  2  &  {\color{ForestGreen}{42\xspace}} / {\color{red}{4\xspace}}  &  80.0  &  18  &  0  &  87.0s  &  0  &  40.1m\\
&  8ac71b7  &  7/6/16 &  210  &  10  &  {\color{ForestGreen}{167\xspace}} / {\color{red}{31\xspace}}  &  40.0  &  20  &  0  &  2.1m  &  \cmark(124)  &  5.6m\\
\midrule
\rowcolor[HTML]{EFEFEF}
&  total  &  \xspace{} &  \xspace{}  &  \xspace{}  &  \xspace{}  &  \xspace{}  &  632  &  1  &  58.1m  &  4422  &  42.0h\\
\midrule
&  average  &  \xspace{} &  \xspace{}  &  \xspace{}  &  \xspace{}  &  \xspace{}  &  63.20  &  0.10  &  5.8m  &  442.20  &  4.2h\\
\midrule
\multirow{11}{*}{\rotvertical{xwiki-commons}}
&  ffc3997  &  7/27/18 &  1081  &  0  &  {\color{ForestGreen}{125\xspace}} / {\color{red}{18\xspace}}  &  21.05  &  1  &  0  &  29.0s  &  0  &  18.0s\\
&  ced2635  &  8/13/18 &  1081  &  1  &  {\color{ForestGreen}{21\xspace}} / {\color{red}{14\xspace}}  &  60.0  &  5  &  0  &  93.0s  &  0  &  2.5h\\
&  10841b1  &  8/1/18 &  1061  &  1  &  {\color{ForestGreen}{107\xspace}} / {\color{red}{19\xspace}}  &  30.0  &  51  &  0  &  5.7m  &  0  &  3.4h\\
&  848c984  &  7/6/18 &  1074  &  1  &  {\color{ForestGreen}{154\xspace}} / {\color{red}{111\xspace}}  &  17.65  &  1  &  0  &  28.0s  &  0  &  18.0s\\
&  adfefec  &  6/27/18 &  1073  &  1  &  {\color{ForestGreen}{17\xspace}} / {\color{red}{14\xspace}}  &  40.0  &  22  &  \cmark(1)  &  76.0s  &  \cmark(3)  &  14.9m\\
&  d3101ae  &  1/18/18 &  1062  &  2  &  {\color{ForestGreen}{71\xspace}} / {\color{red}{9\xspace}}  &  20.0  &  4  &  \cmark(1)  &  72.0s  &  \cmark(31)  &  41.4m\\
&  a0e8b77  &  1/18/18 &  1062  &  2  &  {\color{ForestGreen}{51\xspace}} / {\color{red}{8\xspace}}  &  42.86  &  4  &  \cmark(1)  &  72.0s  &  \cmark(60)  &  42.1m\\
&  78ff099  &  12/19/17 &  1061  &  1  &  {\color{ForestGreen}{16\xspace}} / {\color{red}{0\xspace}}  &  33.33  &  2  &  0  &  68.0s  &  \cmark(4)  &  6.6m\\
&  1b79714  &  11/13/17 &  1060  &  1  &  {\color{ForestGreen}{20\xspace}} / {\color{red}{5\xspace}}  &  60.0  &  22  &  0  &  78.0s  &  0  &  17.9m\\
&  6dc9059  &  10/31/17 &  1060  &  1  &  {\color{ForestGreen}{4\xspace}} / {\color{red}{14\xspace}}  &  88.89  &  22  &  0  &  79.0s  &  0  &  20.5m\\
\midrule
\rowcolor[HTML]{EFEFEF}
&  total  &  \xspace{} &  \xspace{}  &  \xspace{}  &  \xspace{}  &  \xspace{}  &  134  &  3  &  15.7m  &  98  &  8.2h\\
\midrule
&  average  &  \xspace{} &  \xspace{}  &  \xspace{}  &  \xspace{}  &  \xspace{}  &  13.40  &  0.30  &  94.3s  &  9.80  &  49.5m\\
\midrule
\midrule
\rowcolor[HTML]{EFEFEF}
&  total  &  \xspace{} &  \xspace{}  &  \xspace{}  &  \xspace{}  &  \xspace{}  &  3378  &  9(15)  &  2.2h  &  25(6708)  &  100.9h\\
\bottomrule
\end{tabular}
\end{table}

The overall results are reported in \Autoref{tab:overall_result}.
\rev{This table can be read as follows:}
\rev{the first column is the name of the project;}
the second column is the shortened commit id;
the third column is the commit date;
the fourth column column is the total number of test methods executed when building that version of the project;
the fifth and sixth columns are respectively the number of tests modified or added by the commit, and the size of the diff in terms of line additions (in green) and deletions (in red);
the seventh and eighth columns are respectively the diff coverage and the number of tests DCI selected;
the ninth column provides the amplification results for \DCIA, and it is either a \cmark with the number of amplified tests that detect a behavioral change or a \textit{-} if DCI did not succeed in generating a test that detects a change;
the tenth column displays the time spent on the amplification phase;
The eleventh and the twelfth are respectively a \cmark with the number of amplified tests for \DCII  (or - if a change is not detected) for 3 iterations.
\rev{The last row reports the total over the 6 projects.}
\rev{For the tenth and the twelfth columns of the last row, the first number is the number of successes, \ie the number of times \DCI produced at least one amplified test method that detects the behavioral change, for  \DCIA and \DCII respectively.}
\rev{The numbers between brackets correspond to the total number of amplified test methods that \DCI produces in each mode.}

\subsubsection{Characteristics of commits with behavioral changes in the context of continuous integration}
\label{subsubsec:answerq1}

In  this section, we describe the  characteristics of commits introducing behavioral changes in the context of continuous integration
The first five columns in \Autoref{tab:overall_result} describe the characteristics of our benchmark.
The commit dates show that the benchmark is only composed of recent commits.
The most recent is \textsc{gson\#b1fb9ca}, authored 9/22/18, and the oldest is \textsc{commons-io\#5d072ef}, authored 9/10/15.
The number of test methods at the time of the commit shows two aspects of our benchmark:

1) we only have strongly tested projects;

2) we see that the number of tests evolve over time due to test evolution.

Every commit in the benchmark comes with test modifications (new tests or updated tests), and commit sizes are quite diverse.
The three smallest commits are \textsc{commons-io\#703228a}, \textsc{gson\#44cad04} and \textsc{jsoup\#e5210d1} with 6 modifications, and the largest is \textsc{Gson\#45511fd} with 334 modifications.

\begin{figure}
\centering
\includegraphics[width=.95\linewidth]{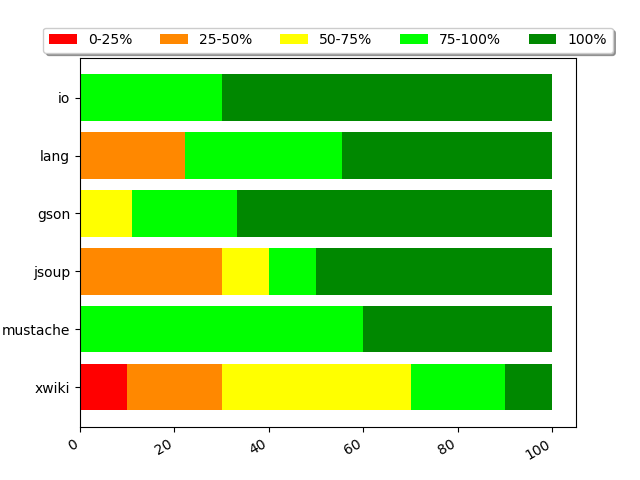}
\caption{Distribution of diff coverage per project of our benchmark.}
\label{fig:histdiffcoverage}
\end{figure}

Finally, on average, commits have 66.11\% coverage. 
The distribution of diff coverage is reported graphically by \Autoref{fig:histdiffcoverage}: 
in commons-io all selected commits have more than 75\% coverage.
In XWiki-Commons, only 50\% of commits have more than 75\% coverage. Overall, 31 / 60 commits have at least 75\% of the changed lines covered.
This validates the correct implementation of our selection criteria that ensures the presence of a test specifying the behavioral change.

Thanks to our selection criteria, we have a curated benchmark of 60 commits with a behavioral change, coming from notable open-source projects, and covering a diversity of commit sizes. The benchmark is publicly available and documented for future research on this topic.

\subsubsection{\rqdetection}
\label{subsubsec:answerq2}

We now focus on the last 4 columns of \Autoref{tab:overall_result}.
For instance, for \textsc{commons-io\#f00d97a} (4$^{th}$ row), DCI$_{AAMPL}$ generated 39 amplified tests that detect the behavioral change. 
For \textsc{commons-io\#81210eb} (8$^{th}$ row), only the \sbampl version of DCI detects the change.
Overall, using only \aampl, DCI generates amplified tests that detect 9 out of 60 behavioral changes.
Meanwhile, using \sbampl only, DCI generates amplified tests that detect 28 out of 60 behavioral changes.

Regarding the number of generated tests.
\DCII generates a large number of test cases, compared to \DCIA only (15 versus 6708, see column ``total'' at the bottom of the table). 
Both \DCIA and \DCII can generate amplified tests, however, since \DCIA does not produce a large amount of test methods, the developers do not have to triage a large set of test cases. 
Also, since \DCIA only adds assertions, the amplified tests are easier to understand than the ones generated by \DCII.

\DCII takes more time than \DCIA (for successful cases 38.7 seconds versus 3.3 hours on average).
The difference comes from the time consumed during the exploration of the input space in the case of \DCII, while \DCIA focuses on the amplification of assertions only, which represents a much smaller space of solutions. 

Overall, DCI successfully generates amplified tests that detect a behavioral change in 42\% of the commits in our benchmark (25 out of 60).
Recall that the 60 commits that we analyze are real changes that fix bugs in complex code bases.
They represent modifications, sometimes deep in the code, that represent challenges with respect to testability~\cite{voas1995software}.
Consequently, the fact that DCI can generate test cases that detect behavioral changes, is considered as an achievement.
The commits for which DCI fails to detect the change can be considered as a target for future research on this topic.

Now, we manually analyze a successful case where DCI detects the behavioral change.
We select commit \textsc{3fadfdd}\footnote{\url{https://github.com/apache/commons-lang/commit/3fadfdd}} from commons-lang, which is succinct enough to be discussed in the paper.
The diff is shown in \Autoref{fig:diff_commons_lang_success}.

\begin{lstlisting}[language=diff,caption=Diff of commit \textsc{3fadfdd} from commons-lang.,label=fig:diff_commons_lang_success]
@@ -2619,7 +2619,7 @@ protected void appendFieldStart(final StringBuffer buffer, final String fieldNam

-       super.appendFieldStart(buffer, FIELD_NAME_QUOTE + fieldName
+       super.appendFieldStart(buffer, FIELD_NAME_QUOTE +
+           StringEscapeUtils.escapeJson(fieldName) + FIELD_NAME_QUOTE);
    }
\end{lstlisting}

The developer added a method call to a method that escapes specials characters in a string. 
The changes come with a new test method that specifies the new behavior. 

\DCI starts the amplification from the \texttt{testNestingPerson} test method defined in \texttt{JsonToStringStyleTest}, \rev{showed in} \Autoref{fig:selected_diff_commons_lang_success}.

\newpage 

\begin{lstlisting}[language=java,caption=Selected test method as a seed to be amplified for commit \textsc{3fadfdd} from commons-lang.,label=fig:selected_diff_commons_lang_success]
@Test
public void testPerson() {
    final Person p = new Person();
    p.name = "Jane Doe";
    p.age = 25;
    p.smoker = true;

    assertEquals(
            "{\"name\":\"Jane Doe\",\"age\":25,\"smoker\":true}",
            new ToStringBuilder(p).append("name", p.name)
                    .append("age", p.age).append("smoker", p.smoker)
                    .toString());
}
\end{lstlisting}

This test is selected for amplification because it triggers the execution of the changed line.

\begin{lstlisting}[language=java,caption=Test generated by DCI that detects the behavioral change of \textsc{3fadfdd} from commons-lang.,label=fig:amplified_commons_lang_success]
@Test(timeout = 10000)
public void testPerson_literalMutationString85602() throws Exception {
    final ToStringStyleTest.Person p = new ToStringStyleTest.Person();
    p.name = "Jane Doe";
    Assert.assertEquals("Jane Doe", p.name);
    p.age = 25;
    p.smoker = true;
    String o_testPerson_literalMutationString85602__6 = new ToStringBuilder(p).append("n/me", p.name).append("age", p.age).append("smoker", p.smoker).toString();
    Assert.assertEquals(
        "{\"n/me\":\"Jane Doe\",\"age\":25,\"smoker\":true}",
        o_testPerson_literalMutationString85602__6
    );
    Assert.assertEquals("Jane Doe", p.name);
}
\end{lstlisting}

We show in \Autoref{fig:amplified_commons_lang_success} the resulting amplified test method.
In this generated test, \DCII applies 2 input transformations: 
1 duplication of method call and 
1 character replacement in an existing String literal.
The latter transformation is the key transformation: DCI replaced an 'a' inside "name" by '/' resulting in "n/me" where "/" is a special character that must be escaped (Line 8). 
Then, DCI generated 11 assertions, based on the modified inputs. 
The amplified test the behavioral change:
in the pre-commit version, the expected value is: 

\texttt{\{\textbackslash"n/me\textbackslash":\textbackslash"Jane Doe\textbackslash",\textbackslash"age\textbackslash":25,\textbackslash"smoker\textbackslash":true\}} 

while in the post-commit version it is 

\texttt{\{\textbackslash"n\textcolor{ForestGreen}{\textbackslash}/me\textbackslash":\textbackslash"Jane Doe\textbackslash",\textbackslash"age\textbackslash":25,\textbackslash"smoker\textbackslash":true\}} (Line 9).
 
\newpage 
 
\begin{mdframed}
Answer to \textbf{RQ1}: Overall, DCI is capable of detecting the behavioral changes for 25/60 commits. \DCII finds behavioral changes in 25/60 commits, while \DCIA finds some in 9/60 commits.
Since \DCII also uses \aampl to generate assertions, all \DCIA's commits are contained in \DCII's. The search-based algorithm of input exploration finds many more behavioral changes, at the cost of execution time.
\end{mdframed}


\subsubsection{\rqiteration}
\label{subsubsec:answerqiteration}

The results are reported in \Autoref{tab:overall_result_iteration}.

\begin{table*}
	\small
	\def\arraystretch{0.6}
	\setlength\tabcolsep{7pt} 
	\caption{Evaluation of the impact of the number of iteration done by \DCII on 60 commits from 6 open-source projects.}
	\label{tab:overall_result_iteration}
	\begin{tabular}{l|c|cc|cc|cc}
		\toprule
		&
id&
$it=1$&
Time&
$it=2$&
Time&
$it=3$&
Time\\
\midrule
\multirow{11}{*}{\rotvertical{commons-io}}
&  c6b8a38  &  0 &  25.0s  &  0  &  62.0s  &  0  &  98.0s\\
&  2736b6f  &  \cmark(1) &  26.1m  &  \cmark(2)  &  44.2m  &  \cmark(12)  &  76.3m\\
&  a4705cc  &  0 &  4.1m  &  0  &  21.1m  &  0  &  38.1m\\
&  f00d97a  &  \cmark(7) &  13.0s  &  \cmark(28)  &  19.0s  &  \cmark(39)  &  27.0s\\
&  3378280  &  \cmark(6) &  15.0s  &  \cmark(10)  &  20.0s  &  \cmark(11)  &  24.0s\\
&  703228a  &  0 &  30.3m  &  0  &  55.1m  &  0  &  71.0m\\
&  a7bd568  &  0 &  28.6m  &  0  &  52.0m  &  0  &  65.2m\\
&  81210eb  &  \cmark(2) &  14.0s  &  \cmark(4)  &  18.0s  &  \cmark(8)  &  23.0s\\
&  57f493a  &  0 &  20.0s  &  0  &  32.0s  &  0  &  54.0s\\
&  5d072ef  &  \cmark(461) &  32.2m  &  \cmark(1014)  &  65.5m  &  \cmark(1538)  &  2.2h\\
\midrule
\rowcolor[HTML]{EFEFEF}
&  total  &  477 &  2.0h  &  1058  &  4.0h  &  1608  &  6.5h\\
\midrule
&  average  &  47.70 &  12.3m  &  105.80  &  24.0m  &  160.80  &  38.8m\\
\midrule
\multirow{11}{*}{\rotvertical{commons-lang}}
&  f56931c  &  0 &  0.0s  &  0  &  3.7m  &  0  &  8.5m\\
&  87937b2  &  0 &  3.5m  &  0  &  10.5m  &  0  &  18.1m\\
&  09ef69c  &  0 &  97.0s  &  0  &  21.0m  &  0  &  98.8m\\
&  3fadfdd  &  \cmark(1) &  2.0m  &  \cmark(1)  &  9.3m  &  \cmark(4)  &  17.2m\\
&  e7d16c2  &  \cmark(3) &  111.0s  &  \cmark(2)  &  8.4m  &  \cmark(2)  &  15.1m\\
&  50ce8c4  &  \cmark(61) &  38.0s  &  \cmark(97)  &  78.0s  &  \cmark(135)  &  2.0m\\
&  2e9f3a8  &  0 &  11.4m  &  0  &  35.0m  &  0  &  66.5m\\
&  c8e61af  &  0 &  16.0s  &  0  &  16.0s  &  0  &  16.0s\\
&  d8ec011  &  0 &  36.0s  &  0  &  68.0s  &  0  &  2.3m\\
&  7d061e3  &  0 &  79.0s  &  0  &  5.8m  &  0  &  11.4m\\
\midrule
\rowcolor[HTML]{EFEFEF}
&  total  &  65 &  23.3m  &  100  &  96.4m  &  141  &  4.0h\\
\midrule
&  average  &  6.50 &  2.3m  &  10.00  &  9.6m  &  14.10  &  24.0m\\
\midrule
\multirow{11}{*}{\rotvertical{gson}}
&  b1fb9ca  &  0 &  14.6m  &  0  &  51.0m  &  0  &  92.5m\\
&  7a9fd59  &  \cmark(7) &  33.0s  &  \cmark(48)  &  73.0s  &  \cmark(108)  &  2.1m\\
&  03a72e7  &  0 &  30.2m  &  0  &  102.3m  &  0  &  3.2h\\
&  74e3711  &  0 &  6.0s  &  0  &  11.0s  &  0  &  16.0s\\
&  ada597e  &  0 &  61.0s  &  0  &  4.9m  &  0  &  8.7m\\
&  a300148  &  0 &  45.2m  &  \cmark(4)  &  2.6h  &  \cmark(6)  &  4.9h\\
&  9a24219  &  0 &  10.8m  &  0  &  28.4m  &  0  &  48.9m\\
&  9e6f2ba  &  0 &  79.0s  &  0  &  4.5m  &  \cmark(2)  &  8.5m\\
&  44cad04  &  \cmark(4) &  21.0s  &  \cmark(21)  &  30.0s  &  \cmark(37)  &  40.0s\\
&  b2c00a3  &  0 &  31.5m  &  0  &  111.8m  &  0  &  3.6h\\
\midrule
\rowcolor[HTML]{EFEFEF}
&  total  &  11 &  2.3h  &  73  &  7.7h  &  153  &  14.4h\\
\midrule
&  average  &  1.10 &  13.6m  &  7.30  &  46.0m  &  15.30  &  86.5m\\
\midrule
\multirow{11}{*}{\rotvertical{jsoup}}
&  426ffe7  &  \cmark(126) &  5.4m  &  \cmark(172)  &  19.2m  &  \cmark(198)  &  33.6m\\
&  a810d2e  &  0 &  90.0s  &  0  &  13.9m  &  0  &  26.6m\\
&  6be19a6  &  0 &  8.1m  &  0  &  39.7m  &  0  &  67.7m\\
&  e38dfd4  &  0 &  117.0s  &  0  &  6.3m  &  0  &  12.5m\\
&  e9feec9  &  0 &  20.0s  &  0  &  50.0s  &  0  &  95.0s\\
&  0f7e0cc  &  \cmark(1) &  2.4h  &  \cmark(7)  &  6.8h  &  \cmark(36)  &  11.8h\\
&  2c4e79b  &  0 &  7.1m  &  0  &  34.1m  &  0  &  4.7h\\
&  e5210d1  &  0 &  45.0s  &  0  &  2.3m  &  0  &  4.9m\\
&  df272b7  &  0 &  43.0s  &  0  &  2.2m  &  0  &  4.6m\\
&  3676b13  &  \cmark(6) &  21.4m  &  \cmark(35)  &  2.9h  &  \cmark(52)  &  6.8h\\
\midrule
\rowcolor[HTML]{EFEFEF}
&  total  &  133 &  3.2h  &  214  &  11.6h  &  286  &  25.8h\\
\midrule
&  average  &  13.30 &  19.4m  &  21.40  &  69.8m  &  28.60  &  2.6h\\
\midrule
\multirow{11}{*}{\rotvertical{mustache.java}}
&  a1197f7  &  \cmark(28) &  5.9h  &  \cmark(124)  &  8.4h  &  \cmark(204)  &  10.1h\\
&  8877027  &  0 &  30.5m  &  0  &  58.4m  &  0  &  100.2m\\
&  d8936b4  &  0 &  3.2m  &  0  &  4.8m  &  0  &  84.2m\\
&  88718bc  &  \cmark(13) &  78.0s  &  \cmark(85)  &  2.5m  &  \cmark(149)  &  3.7m\\
&  339161f  &  \cmark(143) &  115.9m  &  \cmark(699)  &  4.1h  &  \cmark(1312)  &  5.8h\\
&  774ae7a  &  \cmark(18) &  2.7m  &  \cmark(65)  &  4.7m  &  \cmark(124)  &  6.8m\\
&  94847cc  &  \cmark(122) &  5.3h  &  \cmark(580)  &  10.4h  &  \cmark(2509)  &  21.4h\\
&  eca08ca  &  0 &  8.1m  &  0  &  24.3m  &  0  &  41.8m\\
&  6d7225c  &  0 &  7.9m  &  0  &  26.8m  &  0  &  40.1m\\
&  8ac71b7  &  \cmark(2) &  2.7m  &  \cmark(48)  &  3.8m  &  \cmark(124)  &  5.6m\\
\midrule
\rowcolor[HTML]{EFEFEF}
&  total  &  326 &  14.0h  &  1601  &  25.0h  &  4422  &  42.0h\\
\midrule
&  average  &  32.60 &  84.3m  &  160.10  &  2.5h  &  442.20  &  4.2h\\
\midrule
\multirow{11}{*}{\rotvertical{xwiki-commons}}
&  ffc3997  &  0 &  19.0s  &  0  &  18.0s  &  0  &  18.0s\\
&  ced2635  &  0 &  8.0m  &  0  &  31.8m  &  0  &  2.5h\\
&  10841b1  &  0 &  56.2m  &  0  &  2.9h  &  0  &  3.4h\\
&  848c984  &  0 &  18.0s  &  0  &  17.0s  &  0  &  18.0s\\
&  adfefec  &  \cmark(22) &  3.5m  &  \cmark(57)  &  9.9m  &  \cmark(3)  &  14.9m\\
&  d3101ae  &  \cmark(9) &  11.6m  &  \cmark(12)  &  28.2m  &  \cmark(31)  &  41.4m\\
&  a0e8b77  &  \cmark(10) &  12.0m  &  \cmark(17)  &  28.2m  &  \cmark(60)  &  42.1m\\
&  78ff099  &  \cmark(4) &  2.6m  &  \cmark(4)  &  4.6m  &  \cmark(4)  &  6.6m\\
&  1b79714  &  0 &  4.0m  &  0  &  10.7m  &  0  &  17.9m\\
&  6dc9059  &  0 &  4.0m  &  0  &  10.8m  &  0  &  20.5m\\
\midrule
\rowcolor[HTML]{EFEFEF}
&  total  &  45 &  102.8m  &  90  &  4.9h  &  98  &  8.2h\\
\midrule
&  average  &  4.50 &  10.3m  &  9.00  &  29.7m  &  9.80  &  49.5m\\
\midrule
\midrule
&  total  &  23(1057) &  23.7h  &  24(3136)  &  54.9h  &  25(6708)  &  100.9h\\
\bottomrule
	\end{tabular}
\end{table*}

This table can be read as follow:
the first column is the name of the project;
the second column is the commit identifier;
then, the third, fourth, fifth, sixth, seventh and eighth provide the amplification results and execution time for each number of iteration 1, 2, and 3. 
A \cmark indicates  the number of amplified tests that detect a behavioral change and a \textit{-} denotes that DCI did not succeed in generating a test that detects a change.
\rev{The last row reports the total over the 6 projects.}
\rev{For the third, fifth and the seventh columns of the last row, the first number is the number of successes, \ie the number of times that \DCI produced at least one amplified test method that detect the behavioral change, for respectively$iteration=1$, $iteration=2$ and $iteration=3$.}
\rev{The numbers in parentheses are the total number of amplified test methods that \DCI produces with each number of iteration.}

Overall, \DCII generates amplified tests that detect 23, 24, and 25 out of 60 behavioral changes for respectively $iteration=1$, $iteration=2$ and $iteration=3$.
The more iteration \DCII does, the more it explores, the more it generates amplified tests that detect the behavioral changes but the more it takes time also.
When \DCII is used with $iteration=3$, it generates amplified test methods that detect 2 more behavioral changes than when it is used with $iteration=1$ and 1 then when it is used with $iteration=2$.

On average, \DCII generates 18, 53, and 116 amplified tests for respectively $iteration=1$, $iteration=2$ and $iteration=3$.
This number increases by 544\% from $iteration=1$ to $iteration=3$.
This increase is explained by the fact that \DCII explores more with more iteration and thus is able to generate more amplified test methods that detect the behavioral changes.

In average \DCII takes 23, 64, and 105 minutes to perform the amplification for respectively $iteration=1$, $iteration=2$ and $iteration=3$.
This number increases by 356\% from $iteration=1$ to $iteration=3$.

\paragraph{Impact of the randomness}

The number of amplified test methods obtained by the different seeds are reported in \Autoref{tab:overall_result_seeds}.

\begin{table*}
	\small
	\caption{\rev{Number of successes, \ie \DCI produced at least one amplified test method that detects the behavioral changes, for 10 different seeds.}}
	\centering
	\label{tab:overall_result_seeds}
	\begin{tabular}{l|ccccccccccc}
	    \toprule
        Seed	&	ref	&	1	&	2	&	3	&	4	&	5	&	6	&	7	&	8	&	9\\
        \midrule
        \#Success	&	23	&	18	&	17	&	17	&	17	&	19	&	21	&	18	&	21	&	18\\
        \bottomrule
	\end{tabular}
\end{table*}

\rev{This table can be read as follow:}
\rev{the first column is the id of the commit.}
\rev{the second column is the result obtained with the default seed, used during the evaluation for \textbf{RQ$_1$}.}
\rev{the ten following columns are the results obtained for the 10 different seeds.}
\rev{The computed confidence interval is $\left[20.34, 17.66\right]$}
\rev{It means that, from our samples, with probability 0.95, the real value of the number of successes lies in this interval.}

\begin{mdframed}
Answer to \textbf{RQ2}: \DCII detects  23, 24, and 25 behavioral changes out of 60 commits for respectively $iteration=1$, $iteration=2$ and $iteration=3$.
The number of iterations performed by \DCII impacts the number of behavioral changes detected, the number of amplified test methods obtained and the execution time.
\end{mdframed}


\subsubsection{\rqselection}
\label{subsubsec:answerq3}

To answer \textbf{RQ3}, there is no quantitative approach to take, because there is no ground truth data or metrics to optimize. 
Per our protocol described in \Autoref{subsec:protocol}, we answer this question based on manual analysis:
we randomly selected 1 commit per project, and we analyzed the relevance of the selected tests for amplification.

In order to give an intuition of what we consider as a relevant test selection for amplification, let us look at an example. 
If \texttt{TestX} is selected for amplification, following a change to method \texttt{X}, we consider this as relevant. The key is that DCI will generate an amplified test \texttt{TestX'} that is a variant of \texttt{TestX}, and, consequently, the developer will directly get the intention of the new test \texttt{TestX'} and what behavioral change it detects.

\textsc{Commons-io\#c6b8a38}\footnote{\url{https://github.com/apache/commons-io/commit/c6b8a38}}: our test selection returns 3 test methods: \texttt{testContentEquals}, \texttt{testCopyURLToFileWithTimeout} and \texttt{testCopyURLToFile} from the same test class: \texttt{FileUtilsTestCase}.
The considered commit modifies the method \texttt{copyToFile} from \texttt{FileUtils}. 
Two test methods out of 3 (\texttt{testCopyURLToFileWithTimeout} and \texttt{testCopyURLToFile}) have an intention related to the changed file. 
The selection is thus considered relevant.

\textsc{Commons-lang\#f56931c}\footnote{\url{https://github.com/apache/commons-lang/commit/f56931c}}: our test selection returns 39 test methods from 5 test classes: \texttt{FastDateFormat\_ParserTest}, \texttt{FastDateParserTest}, \texttt{DateUtilsTest}, \texttt{FastDateParser\_TimeZoneStrategyTest} and \texttt{FastDateParser\_MoreOrLessTest}.
This commit modifies the behavior of two methods: \texttt{simpleQuote} and \texttt{setCalendar} of class \texttt{FastDateParser}.
Our manual analysis reveals two intentions:
1) test behaviors related to parsing, 
1) test behaviors related to dates.
While this is meaningful, a set of 39 methods is not a focused selection.
It is considered as an half-success.

 \textsc{Gson\#9e6f2ba}\footnote{\url{https://github.com/google/gson/commit/9e6f2ba}}: our test selection returns 9 test methods from 5 different test classes.
 Three out of those five classes \texttt{JsonElementReaderTest}, \texttt{JsonReaderPathTest} and \texttt{JsonParserTest} relate to the class modified in the commit(\texttt{JsonTreeReader}).
The selection is thus considered relevant but unfocused.

 \textsc{Jsoup\#e9feec9}\footnote{\url{https://github.com/jhy/jsoup/commit/e9feec9}}, our test selection returns the 4 test methods defined in the \texttt{XmlTreeBuilderTest} class : \texttt{caseSensitiveDeclaration}, \texttt{handlesXmlDeclarationAsDeclaration}, \texttt{testDetectCharsetEncodingDeclaration} and \texttt{testParseDeclarationAttributes}.
 The commit modifies the behavior of the class \texttt{XmlTreeBuilder}.
Here, the test selection is relevant.
Actually, the ground-truth, manually written test added in the commit is also in the \texttt{XmlTreeBuilderTest} class.
If DCI proposes a new test there to capture the behavioral change, the developer will understand its relevance and its relation to the change.

\textsc{Mustache.java\#88718bc}\footnote{\url{https://github.com/spullara/mustache.java/commit/88718bc}}, our test selection returns the \texttt{testInvalidDelimiters} test method defined in the \texttt{com.github.mustachejava.InterpreterTest} test class.
The commit improves an error message when an invalid delimiter is used.
Here, the test selection is relevant since it selected \texttt{testInvalidDelimiters} which is the dedicated test to the usage of the test invalid delimiters.
This ground-truth test method is also in the test class \texttt{com.github.mustachejava.InterpreterTest}.

\textsc{Xwiki-commons\#848c984}\footnote{\url{https://github.com/xwiki/xwiki-commons/commit/848c984}} our test selection returns a single test method \texttt{createReference} from test class \texttt{XWikiDocumentTest}.
The main modification of this commit is on class \texttt{XWikiDocument}.
Since \texttt{XWikiDocumentTest} is the test class dedicated to \texttt{XWikiDocument}, this is considered as a success.

\begin{mdframed}
Answer to \textbf{RQ3}: 
In 4 out of the 6 manually analyzed cases, the tests selected to be amplified are semantically related to the modified application code. 
In the 2 remaining cases, DCI selects tests whose intention is semantically relevant to the change, but also tests that are not.
DCI's test selection provides developers with important and targeted context to better understand the behavioral change at hand.
\end{mdframed}

\subsubsection{\rqhuman}
\label{subsubsec:answerq4}

When DCI generates an amplified test method that detects the behavioral change, we can compare it to the ground truth version (the test added in the commit) to see whether it captures the same behavioral change.
For each project, we select 1 successful application of DCI, and we compare the DCI test against the human test.\footnote{For a side-by-side comparison, see \url{https://danglotb.github.io/resources/dci/index.html}}
If they capture the same behavioral change, it means they have the same intention and we consider the amplification a success.


\textsc{commons-io\#81210eb}\footnote{\url{https://github.com/apache/commons-io/commit/81210eb}}: This commit modifies the behavior of the \texttt{read()} method in \texttt{BoundedReader}.
\Autoref{fig:ampl_commons-io} shows the test generated by \DCII.
This test is amplified from the existing \texttt{readMulti} test, which indicates that the intention is to test the read functionality.
The first line of the test is the construction of a \texttt{BoundedReader} object (Line 3) which is also the class modified by the commit.
\DCII modified the second parameter of the constructor call (transformed $3$ into a $0$) and generated two assertions (only 1 is shown).
The first assertion, associated to the new test input, captures the behavioral difference.
Overall, this can be considered as a successful amplication.



\begin{lstlisting}[language=java,caption=Test generated by \DCII that detects the behavioral change introduced by commit \textsc{81210eb} in commons-io.,label=fig:ampl_commons-io]
@Test(timeout = 10000)
public void readMulti_literalMutationNumber3() {
    BoundedReader mr = new BoundedReader(sr, 0);
    char[] cbuf = new char[4];
    for (int i = 0; i < (cbuf.length); i++) {
        cbuf[i] = 'X';
    }
    final int read = mr.read(cbuf, 0, 4);
    Assert.assertEquals(0, ((int) (read)));
}        
\end{lstlisting}

Now, let us look at the human test contained in the commit, shown in \Autoref{fig:diff_commons-io}.
It captures the behavioral change with the timeout (the test timeouts on the pre-commit version and goes fast enough on the post-commit version). 
Furthermore, it only indirectly calls the changed method through a call to \texttt{readLine}.

In this case, the DCI test can be considered better than the developer test because
1) it relies on assertions and not on timeouts, and
2) it directly calls the changed method (\texttt{read}) instead of indirectly. 



\begin{lstlisting}[language=java,caption=Developer test for commit \textsc{81210eb} of commons-io.,label=fig:diff_commons-io]
@Test(timeout = 5000)
public void testReadBytesEOF() {
   BoundedReader mr = new BoundedReader( sr, 3 );
   BufferedReader br = new BufferedReader( mr );
   br.readLine();
   br.readLine();
}
\end{lstlisting}

\textsc{commons-lang\#e7d16c2}\footnote{\url{https://github.com/apache/commons-lang/commit/e7d16c2}}: this commit escapes special characters before adding them to a \texttt{StringBuffer}.
\Autoref{fig:ampl_commons-lang} shows the amplified test method obtained by \DCII.
The assertion at the bottom of the excerpt is the one that detects the behavioral change.
This assertion compares the content of the \texttt{StringBuilder} against an expected string.
In the pre-commit version, no special character is escaped, \eg '\textbackslash |'.
In the post-commit version, the DCI test fails since the code now escapes the special character \textbackslash.



\newpage 

\begin{lstlisting}[language=java,caption=Test generated by \DCII that detects the behavioral change of \textsc{e7d16c2} in commons-lang.,label=fig:ampl_commons-lang]
@Test(timeout = 10000)
public void testAppendSuper_literalMutationString64() {
    String o_testAppendSuper_literalMutationString64__15 = 
        new ToStringBuilder(base)
            .appendSuper((((("Integer@8888[" + (System.lineSeparator())) + "  null") 
                + (System.lineSeparator())) + "]"))
            .append("a", "b0/|]")
            .toString();
    Assert.assertEquals("{\"a\":\"b0/|]\"}", o_testAppendSuper_literalMutationString64__15);
}
\end{lstlisting}

Let's have a look ar the human test method shown in \Autoref{fig:diff_commons-lang}.
Here, the developer specified the new escaping mechanism with 5 different inputs.
The main difference between the human test and the amplified test is that the human test is more readable and uses 5 different inputs.
However, the amplified test generated by DCI is valid since it detects the behavioral change correctly.



\begin{lstlisting}[language=java,caption=Developer test for \textsc{e7d16c2} of commons-lang.,label=fig:diff_commons-lang]
@Test
public void testLANG1395() {
    assertEquals("{\"name\":\"value\"}",
        new ToStringBuilder(base).append("name","value").toString());
    assertEquals("{\"name\":\"\"}",
        new ToStringBuilder(base).append("name","").toString());
    assertEquals("{\"name\":\"\\\"\"}",
        new ToStringBuilder(base).append("name",'"').toString());
    assertEquals("{\"name\":\"\\\\\"}",
        new ToStringBuilder(base).append("name",'\\').toString());
    assertEquals("{\"name\":\"Let's \\\"quote\\\" this\"}",
        new ToStringBuilder(base).append("name","Let's \"quote\" this").toString());
}
\end{lstlisting}

\textsc{gson\#44cad04}\footnote{\url{https://github.com/google/gson/commit/44cad04}}: This commit allows Gson to deserialize a number represented as a string.
\Autoref{fig:ampl_gson} shows the relevant part of the test generated by DCI$_{SBAMPL}$, based on \texttt{testNumberDeserialization} of \texttt{PrimitiveTest} as a seed.
First, we see that the test selected as a seed is indeed related to the change in the deserialization feature.
The DCI test detects the behavioral change at lines 3 and 4.
On the pre-commit version, line 4 throws a \texttt{JsonSyntaxException}.
On the post-commit version, line 5 throws a \texttt{NumberFormatException}.
In other words, the behavioral change is detected by a different exception (different type and not thrown at the same line).
\footnote{Interestingly, the number is parsed lazily, only when needed. Consequently, the exception is thrown when invoking the \texttt{longValue()} method and not when invoking \texttt{parse()}}.



\newpage 

\begin{lstlisting}[language=java,caption=Test generated by DCI that detects the behavioral change of commit \textsc{44cad04} in Gson.,label=fig:ampl_gson]
public void testNumberDeserialization_literalMutationString8_failAssert0() throws Exception {
    try {
        String json = "dhs";
        actual = gson.fromJson(json, Number.class);
        actual.longValue();
        junit.framework.TestCase.fail(
        "testNumberDeserialization_literalMutationString8 should have thrown JsonSyntaxException");
    } catch (JsonSyntaxException expected) {
        TestCase.assertEquals("Expecting number, got: STRING", expected.getMessage());
    }
}
\end{lstlisting}

We compare it against the developer-written ground-truth method, shown in \Autoref{fig:diff_gson}. 
This short test verifies that the program handles a number-as-string correctly.
For this example, the DCI test does indeed detect the behavioral change, but in an indirect way.
On the contrary, the developer test is shorter and directly targets the changed behavior, which is better.



\begin{lstlisting}[language=java,
caption=Provided test by the developer for \textsc{44cad04} of Gson.,
label=fig:diff_gson
]
public void testNumberAsStringDeserialization() {
    Number value = gson.fromJson("\"18\"", Number.class);
    assertEquals(18, value.intValue());
}
\end{lstlisting}

\textsc{jsoup\#3676b13}\footnote{\url{https://github.com/jhy/jsoup/commit/3676b13}}: This change is a pull request (\ie a set of commits) and introduces 5 new behavioral changes. There are two improvements: skip the first new lines in pre tags and support deflate encoding, and three bug fixes: throw exception when parsing some urls, add spacing when output text, and no collapsing of attribute with empty values.
\Autoref{fig:ampl_jsoup} shows an amplified test obtained using \DCII.
This amplified test has 15 assertions and a duplication of method call.
Thanks to this duplication and assertion generated on the \texttt{toString()} method, this test is able to capture the behavioral change introduced by the commit.



\begin{lstlisting}[language=java,
caption=Test generated by \DCII that detects the behavioral change of \textsc{3676b13} of Jsoup.,
label=fig:ampl_jsoup
]
@Test(timeout = 10000)
public void parsesBooleanAttributes_add4942() {
    String html = "<a normal=\"123\" boolean empty=\"\"></a>";
    Element el = Jsoup.parse(html).select("a").first();
    List<Attribute> attributes = el.attributes().asList();
    Attribute o_parsesBooleanAttributes_add4942__15 = 
    attributes.get(1);
    Assert.assertEquals("boolean=\"\"", 
        ((BooleanAttribute) (o_parsesBooleanAttributes_add4942__15)).toString());
}
\end{lstlisting}

As before, we compare it to the developer's test. 
The developer uses the \texttt{Element} and \texttt{outerHtml()} methods rather than \texttt{Attribute} and \texttt{toString()}.
However, the method \texttt{outerHtml()} in \texttt{Element} will call the \texttt{toString()} method of \texttt{Attribute}.
For this behavioral change, it concerns the \texttt{Attribute} and not the \texttt{Element}.
So, the amplified test is arguably better, since it is closer to the change than the developer's test.
But, \DCII generates amplified tests that detect 2 of 5 behavioral changes: adding spacing when output text and no collapsing of attribute with empty values only, so regarding the quantity of changes, the human tests are more complete.



\begin{lstlisting}[language=java,
caption=Provided test by the developer for \textsc{3676b13} of Jsoup.,
label=fig:diff_jsoup
]
@Test
public void booleanAttributeOutput() {
    Document doc = Jsoup.parse("<img src=foo noshade='' nohref async=async autofocus=false>");
    Element img = doc.selectFirst("img");

    assertEquals("<img src=\"foo\" noshade nohref async autofocus=\"false\">", img.outerHtml());
}
\end{lstlisting}

\textsc{Mustache.java\#774ae7a}\footnote{\url{https://github.com/spullara/mustache.java/commit/774ae7a}}: This commit fixes an issue with the usage of a dot in a  relative path on Window in the method \texttt{getReader} of class \texttt{ClasspathResolver}.
The test method \texttt{getReaderNullRootDoesNotFindFileWithAbsolutePath} has been used as seed by \DCI. 
It modifies the existing string literal with another string used somewhere else in the test class and generates 3 new assertions.
The behavioral change is detected thanks to the modified strings: it produces the right test case containing a space.



\begin{lstlisting}[language=java,
caption=Test generated by \DCII that detects the behavioral change of \textsc{774ae7a} of Mustache.java.,
label=fig:ampl_mustache
]
@Test(timeout = 10000)
public void getReaderNullRootDoesNotFindFileWithAbsolutePath_litStr4() {
    ClasspathResolver underTest = new ClasspathResolver();
    Reader reader = underTest.getReader(" does not exist");
        Assert.assertNull(reader);
    Matcher<Object> 
        o_getReaderNullRootDoesNotFindFileWithAbsolutePath_litStr4__5 =
            Is.is(CoreMatchers.nullValue());
    Assert.assertEquals("is null", 
        ((Is) (o_getReaderNullRootDoesNotFindFileWithAbsolutePath_litStr4__5))
            .toString());
    Assert.assertNull(reader);
}
\end{lstlisting}

The developer proposed two tests that verify that the object reader is not null when getting it with dots in the path.
There are shown in \Autoref{fig:diff_mustache}.
These tests invoke the method \texttt{getReader} which is the modified method in the commit.
The difference is that the \DCII's amplified test method provides a non longer valid input for the method \texttt{getReader}.
However, providing such inputs produce errors afterward which signal the behavioral change.
In this case, the amplified test is complementary to the human test since it verifies that the wrong inputs are no longer supported and that the system immediately throws an error.



\begin{lstlisting}[language=java,
caption=Developer test for \textsc{774ae7a} of Mustache.java.,
label=fig:diff_mustache
]
@Test
public void getReaderWithRootAndResourceHasDoubleDotRelativePath() throws Exception {
    ClasspathResolver underTest = new ClasspathResolver("templates");
    Reader reader = underTest.getReader("absolute/../absolute_partials_template.html");
    assertThat(reader, is(notNullValue()));
}

@Test
public void getReaderWithRootAndResourceHasDotRelativePath() throws Exception {
    ClasspathResolver underTest = new ClasspathResolver("templates");
    Reader reader = underTest.getReader("absolute/./nested_partials_sub.html");
    assertThat(reader, is(notNullValue()));
}
\end{lstlisting}

\textsc{xwiki-commons\#d3101ae}\footnote{\url{https://github.com/xwiki/xwiki-commons/commit/d3101ae}}: This commit fixes a bug in the \texttt{merge} method of class \texttt{DefaultDiffManager}.
\Autoref{fig:ampl_xwiki} shows the amplified test method obtained by \DCIA.
DCI used \texttt{testMergeCharList} as a seed for the amplification process, and generates 549 new assertions.
Among them, 1 assertion captures the behavioral change between the two versions of the program: 
``assertEquals(0, result.getLog().getLogs(LogLevel.ERROR).size());''.
The behavioral change that is detected is the presence of a new logging statement in the diff. After verification, there is indeed such a behavioral change in the diff, with the addition of a call to ``logConflict'' in the newly handled case.



\begin{lstlisting}[language=java,
caption=Test generated by \DCIA that detects the behavioral change of \textsc{d3101ae} of XWiki.,
label=fig:ampl_xwiki
]
@Test(timeout = 10000)
public void testMergeCharList() throws Exception {
    MergeResult<Character> result;
    result = this.mocker.getComponentUnderTest().merge(AmplDefaultDiffManagerTest.toCharacters("a"), AmplDefaultDiffManagerTest.toCharacters(""), AmplDefaultDiffManagerTest.toCharacters("b"), null);
    int o_testMergeCharList__9 = result.getLog().getLogs(LogLevel.ERROR).size();
    Assert.assertEquals(1, ((int) (o_testMergeCharList__9)));
    List<Character> o_testMergeCharList__12 = AmplDefaultDiffManagerTest.toCharacters("b");
    Assert.assertTrue(o_testMergeCharList__12.contains('b'));
    result.getMerged();
    result = this.mocker.getComponentUnderTest().merge(AmplDefaultDiffManagerTest.toCharacters("bc"), AmplDefaultDiffManagerTest.toCharacters("abc"), AmplDefaultDiffManagerTest.toCharacters("bc"), null);
    int o_testMergeCharList__21 = result.getLog().getLogs(LogLevel.ERROR).size();
    Assert.assertEquals(0, ((int) (o_testMergeCharList__21)));
}
\end{lstlisting}

The developer's test is shown in \Autoref{fig:diff_xwiki}.
This test method directly calls method \texttt{merge}, which is the method that has been changed. 
What is striking in this test is the level of clarity: the variable names, the explanatory comments and even the vertical space formatting are impossible to achieve with \DCIA and makes the human test clearly of better quality but also longer to write. 
Yet, \DCIA's amplified tests capture a behavioral change that was not specified in the human test.
In this case, amplified tests can be complementary.



\begin{lstlisting}[language=java,
caption=Developer test for \textsc{d3101ae} of XWiki.,
label=fig:diff_xwiki
]
@Test
public void testMergeWhenUserHasChangedAllContent() throws Exception
{
    MergeResult<String> result;

    // Test 1: All content has changed between previous and current
    result = mocker.getComponentUnderTest().merge(Arrays.asList("Line 1", "Line 2", "Line 3"),
    Arrays.asList("Line 1", "Line 2 modified", "Line 3", "Line 4 Added"),
    Arrays.asList("New content", "That is completely different"), null);

    Assert.assertEquals(Arrays.asList("New content", "That is completely different"), result.getMerged());

    // Test 2: All content has been deleted 
    // between previous and current
    result = mocker.getComponentUnderTest().merge(Arrays.asList("Line 1", "Line 2", "Line 3"),
    Arrays.asList("Line 1", "Line 2 modified", "Line 3", "Line 4 Added"),
    Collections.emptyList(), null);

    Assert.assertEquals(Collections.emptyList(), result.getMerged());
}
\end{lstlisting}

\newpage 

\begin{mdframed}
Answer to \textbf{RQ4}: 
In 3 out of 6 cases, the DCI test is complementary to the human test.
In 1 case, the DCI test can be considered better than the human test.
In 2 cases, the human test is better than the DCI test.
Even though human tests can be better, DCI can be complementary and catch missed cases, and provide added-value when developers do not have the time to add a test.
\end{mdframed}


\section{Discussion about the scope of DCI}
\label{sec:limitation}

In this section, we overview the current scope of DCI and the key challenges that limit DCI.

\textbf{Focused applicability}
\rev{From our benchmark, we see that \DCI is applicable to a limited proportion of commits: on average 15.29\% of the commits analyzed.}
\rev{This low proportion is the first limit of \DCI usage.}
\rev{However, once \DCI is setup, it is fully automated, there is no manual overhead.}
\rev{Even if \DCI is not used at each commit, it costs little more.}

\textbf{Adoption}
\rev{Our evaluation showed that \DCI is able to obtain amplified test methods that detect behavioral changes.}
\rev{But, it does not provide any evidence on the fact that developers would exploit such test methods.}
\rev{However, from our past evaluation~\cite{dspot-emse}, we know that software developers value amplified test methods.}
\rev{This provides strong evidence of the potential adoption of \DCI.}

\textbf{Performance}
From our experiments, we see that the time to complete the amplification is the main limitation of DCI. For example \DCI took almost 5 hours on
\textsc{jsoup\#2c4e79b},  with no result.
For the sake of our experimentation, we choose to use a pre-defined number of iterations to bound the exploration.
In practice, we recommend to set a time budget (\eg at most one hour per pull-request).

\textbf{Importance of test seeds}
By construction, DCI's effectiveness is correlated to the test methods used as seeds.
For example, see the row of \texttt{commons-lang\#c8e61af} in \Autoref{tab:overall_result_iteration}, where one can observe that whatever the number of iterations, DCI takes the same time to complete the amplification.
The reason is that the seed tests are only composed of assertions statements.
Such tests are bad seeds for DCI, and they prevent any good input amplification.
\rev{Also, \DCI requires to have at least one test method that executes the code changes.}
\rev{If the project is poorly tested and does not have any test method that execute the code changes, \DCI cannot be applied.}

\textbf{False positives}
The risk of false positives is a potential  limitation of our approach.
A false positive would be an amplified test method that passes or fails on both versions, which means that the amplified test method does not detect the behavioral difference between both versions.
We manually analyzed 6 commits and none of them are false positives.
This increases our confidence that DCI produces a limited number of such confusing test methods.

\section{Threats to validity}
\label{sec:threats}

An internal threat is the potential bugs in the implementation of DCI.
However, we heavily tested our prototype with JUnit test cases to mitigate this threat.

In our benchmark, there are 60 commits. 
Our result may be not be generalizable to all programs. But we carefully selected real and diverse applications from GitHub, all having a strong test suite.
We believe that the benchmark reflects real programs, and we have good confidence in the results.

Last but not least, there is a potential flakiness to generated test methods.
However we take care that our approach does not produce flaky test methods, and we make sure to observe a stable and different state of the program between different executions. 
To do this, we execute each amplified test 3 times in order to check weather or not there are stable.
If the outcome of at least one execution is different than the others, we discard the amplified test.

\rev{Our experiments are stochastic, and randomness is a threat accordingly.}
\rev{To mitigate this threat, we have computed a confidence interval that estimates the number of successes that \DCI would obtain.}


\section{Related Work}
\label{sec:related_work}

\subsection{Commit-based test generation}

Person \etal~\cite{dse} present differential symbolic execution (DSE). DSE combines symbolic execution and a new approximation technique to highlight behavioral changes.
They use symbolic execution summary to find equivalences and difference and generate a set of inputs that trigger different behavior.
This is done in three steps: 1) they execute both versions of the modified method; 2) they find equivalences and differences, thanks to the analysis of symbolic execution summary;  3) they generate a set of inputs that trigger the different behaviors in both versions.
The main difference with our work is that they have the strong assumption to have a program whose semantics is fully handled by the symbolic execution engine. 
In the context of Java, to our knowledge, no symbolic execution engine works on arbitrary Java program.
Symbolic execution engines do not scale to the size and complexity of the programs we targeted.
On the contrary, our approach, being more lightweight, is meant to work on all Java programs.

Marinescu and Cadar~\cite{marinescu2013katch} present Katch, a system that aims at covering the code included in a patch.
This approach first determine[17.66 ; 20.34s the differences of a program and its previous version.
It targets modified and not executed by the existing test suite lines.
Then, it selects the closest input to each target from existing tests using a static minimum distance over the control flow graph.
The proposal is evaluated on Unix tools. 
They examine patches from a period of 3 years. In average, they automatically increase coverage from 35\% to 52\% with respect to the manually written test suite.
Contrary to our work, they only aim at increasing the coverage, not at detecting behavioral changes.

A posterior work of the same group~\cite{palikareva2016shadow,Kuchta:2018:SSE:3276753.3208952} focuses on finding test inputs that execute different behaviors in two program versions. 
They devise a technique, named ShaddowKlee, built on top of Klee \cite{klee}. 
They require the code to be annotated at changed places. 
Then they select from the test suite those test cases that cover the changed code. The unified program is used in a two stage dynamic symbolic execution guided by the selected test cases.
They first look for branch points where the conditions are evaluated in both program versions.
Then, a constraint solver generates new test inputs for divergent scenarios. The program versions are then normally executed with the generated inputs and the result is validated to check the presence of a bug or of an intended difference.
The evaluation of the proposed method is based on the CoREBench~\cite{bohme2014corebench} data set that contains documented regression bugs of the GNU Coreutils program suite. 

Noller \etal~\cite{jpfshadow} aim at detecting regression bugs. 
They apply shadow symbolic execution, originally from Palikevera~\cite{dse,palikareva2016shadow} that has been discussed in the previous paragraph, on Java programs. 
Their approach has been implemented as an extension of Java Path Finder Symbolic (jpf-symbc)\cite{jpfsymb}, named jpf-shadow. 
Shadow symbolic execution generate test inputs that trigger the new program behavior. 
They use a merged version of both version of the same program, i.e. the previous version, so called old, and the changed version, called new
This is done by instrumenting the code with method calls ``change()''.
The method change() takes two inputs: the old statement and the new one.[17.66 ; 20.34
Then, a first step collects divergence points, i.e. conditional where the old version and the new version do not take the same branch.
On small examples, they show that jpf-shadow generates less unit test cases yet cover the same number of path. 
Jpf-shadow only aims at covering the changes and not at detecting the behavioral change with an assertion.

Menarini \etal~\cite{semantics:code:review} proposes a tool, GETTY, based on invariants mined by Daikon.
GETTY provides to code reviewers a summary of the behavioral changes, based on the difference of invariants for various combinations of programs and test suites.
They evaluate GETTY on 6 open source project, and showed that their behavioral change summaries can detect bugs earlier than with normal code review.
While they provide a summary, DCI provides a concrete test method with assertions that detect the behavioral changes. 

Lahiri \etal~\cite{differential-assertion-checking} propose differential assertion checking (DAC): checking two versions of a program with respect to a set of assertions. DAC is based on filtering false alarms of verification analysis. They evaluate DAC on a set of small example.
The main difference is that DAC requires to manually write specifications, while DCI is completely automated with normal code as input.

Yang \etal~\cite{Yang:2014:PDI:2568225.2568319} introduce IProperty, a  way to annotate  correctness properties of programs. They evaluate their approach on the triangle problem.
The key novelty of our work is to perform an evaluation on real commits from large scale open source software.

Campos \etal~\cite{Campos:2014:CTG:2642937.2643002} extended EvoSuite to adapt test generation techniques to continuous integration.
Their contribution is the design of a time budget allocation strategy: it allocates more time budget to specific classes that are involved in the changes.
They evaluated their approach on 10 projects from the SF100 corpus, on 8 of the most popular open-source projects from GitHub, and on 5 industrial projects.
They limit their evaluation to the 100 last consecutive commits.
They observe an increase of +58\% branch coverage, +69\% thrown undeclared exceptions, while reducing the time consumption by up to 83\% compared to the baseline.
The major difference compared to our approach, they do not aim at specifically obtaining test methods that detect the behavioral changes but rather obtain better branch coverage and detect undeclared exceptions. They also do not generate any assertions.
However, from the point of view practitioners, integrating a time budget strategy into DCI would increase its usability, practicability and potential adoption.

\subsection{Behavioral change detection}

Evans \etal \cite{evans2007differential} devise the differential testing.
This approach aims at alleviating the test repair
problem and detects more changes than regression testing alone.
They use an automated characterization test
generator (ACTG) to generate test suite for both version of the program.
They then categorizes the tests of these 2 test suites into 3 groups:
1) $T_{preserved}$ which are the tests that pass on the both versions;
2) $T_{regressed}$ which are the tests that pass on the previous version but not on the new one;
3) $T_{progressed}$ which are the tests that pass on the new version but not on the previous one;
Then, they define also $T_{different}$ which is the union of both $T_{regressed}$ and $T_{progressed}$.
The approach is to execute $T_{different}$ on both versions and observe progressed and regressed behaviors.
They evaluate their approach on a small use case from the SIR dataset on 38 diffrent changes, for version of the program.
They showed that their approach detects 21\%, 34\%, and 21\% more behavior changes than regression testing alone for respectively version 1, version 2 and version 3.
In DCI, the amplified test methods obtained would lie into the $T_{regressed}$ group.
However, we could also amplified test methods using the new version of the program and obtain a $T_{progressed}$.
We would obtain a $T_{different}$ of amplified test methods and it might improve the performance of DCI.
About the evaluation, we run experimentation of 60 commits which the double than their dataset, and on real projects and real commits from \gh.

Wei Jin \etal \cite{automated-behavioral-regression-testing} propose BEhavioral Regression
Testing BERT.
BERT aims at assisting practitioners during development to identify potential regression.
It has been implemented as a plugin for the IDE Eclipse.
Each time a developer make a change in their code base and Eclipse compiles, BERT is triggered.
BERT works in 3 phases:
1) it analyzes what are the classes modified and runs a test generation tools, such as Randoop, to create new test input for these classes.
2) it executes the generated tests on both version of the program and collect multiples values such as the values of the fields of objects, the returned values by methods, etc.
3) it produces a report containing all the differences of behaviors based on the collected values.
Then the developer used this report to decide whether or not the changes are correct.
They evaluated BERT on a small and artificial project, showing that about 60\% of the automatically generated test inputs were able to reveal the behavioral difference that indicates the regression fault
In addition to this proof-of-concept, they evaluated in on JODA-time, which is a mature and widely used library.
They evaluated on 54 pairs of versions.
They reported 36 behavioral differences.
However, they could establish only for one of them was a regression fault.
There are two major differences with DCI:
1) DCI works at commit level and not to the class changes level.
2) DCI produces real and actionable test methods.

Taneja \etal \cite{Taneja:2008:DAR:1642931.1642986} present DiffGen, a tool that generate regression tests for two version of the same class.
Their approach works as follow:
First, they detect the changes between the two version of the class.
It is done using the textual representation and at method level.
Second, they generate what they call a test driver, which is a class that contains a method for each modified method.
These methods takes as input an instance of the old version of the class and the inputs required by the modified method.
They also make all the field public to compare their values between the old version and the new one.
These comparison have the form of branches.
The intuition is if the test generator engine is able to cover these branches, it will reveal the behavioral differences.
Third, they generate test using a test generator and the test driver.
Eventually, they execute the generated tests to see whether or not there is a behavioral difference.
They evaluated DiffGen on 8 artificial classes from the state of the art.
They compared the mutation score of their generated test suite to an existing method from the state of the art.
They showed that that DiffGen has an Improvement Factor IF2 varying from 23.4\% to 100\% for all the subjects.
They also performed an evaluation on larger subjects from the SIR dataset.
They detected 5 more faults than the state of the art.
DiffGen must modify the application code to be efficient while DCI does not required any modification of it.
Thus, is makes generated tests by DiffGen unused by developers since they must expose all the fields of their classes.

Madeiral \etal \cite{Madeiral2019} built a benchmark of bugs for evaluating automatic program repair tools.
This benchmark has been built using behavioral change detection such as we d do in this paper.
However, this benchmark includes a different kind of behavioral change: bug fixes.
Also, they have different criteria to select the commits than ours, and their procedure is similar in different ways.
Their approach used continuous integration to build automatically and enrich their benchmark, and it would be fruitful to automate our process as well.

\subsection{Test amplification}

Yoo \etal \cite{Yoo:2012:TDR:2237756.2237758} devise Test Data Regeneration(TDR). They use hill climbing on existing test data (set of input) that meets a test objective (\eg cover all branch of a function).
The algorithm is based on \emph{neighborhood} and a \emph{fitness} functions as the classical hill climbing algorithm.
The key difference with DCI is that they at fulfilling a test criterion, such as branch coverage, while we aim at obtaining test methods that detect the behavioral changes.

It can be noted that several test generation techniques start from a seed and evolve it to produce a good test suite. This is the case for techniques such as concolic test generation \cite{godefroid2005dart}, search-based test generation \cite{fraser2012seed}, or random  test generation \cite{groce2007randomized}.
The key novelty of DCI relies in the very nature of the tests we used as seed.
DCI uses complete program, which creates objects, manipulates the state of these objects, calls methods on these objects and asserts properties on their behavior. That is to say real and complex object-oriented tests as seed

\subsection{Continuous Integration}

Hilton \etal~\cite{Hilton:2016:UsageCI} conduct a study on the usage, costs and benefits of CI.
To do this, they use three sources:  open-source code, builds from Travis, and they surveyed 442 engineers.
Their studies show that the usage of CI services such as Travis is widely used and became the trend.
The fact that CI is widely used shows that relevance of behavioral change detection.

Zampetti \etal~\cite{static:analysis:in:ci} investigate the usage of Automated Static Code Analysis Tools (ASCAT) in CI.
There investigation is done on 20 projects on \gh.
According to their findings, coding guideline checkers are the most used static analysis tools in CI.
This paper shows that dynamic analysis, such as DCI, is the next step for getting more added-value from CI.   

Spieker \etal~\cite{Spieker:RL:selection} elaborate a new approach for test case prioritization in continuous integration based on reinforcement learning.
Test case prioritization is different from behavioral change detection.

Waller \etal~\cite{Waller:2015:IPB:2735399.2735416} study the portability of performance tests in continuous integration.
They show little variations of performance tests between runs (every night) and claim that the performance tests must be integrated in the CI, early as possible in the development of Software.
Performance testing is also one kind of dynamic analysis for the CI, but different in nature from behavioral change detection.

\section{Conclusion}
\label{sec:conclusion}

In this paper, we have studied the problem of behavioral change detection for continuous integration. 
We have proposed a novel technique called \DCI, which uses assertion generation and search-based transformation of test code to generate tests that automatically detect behavioral changes in commits.
\rev{We analyzed 1576 commits from 6 projects.}
\rev{On average, our approach is applicable to 15.29\% of commits per-project.}
\rev{We built a curated set of 60 commits coming from real-world, large open-source Java projects to evaluate our technique.}
\rev{We show that our approach is able to detect the behavioral differences of 25 of the 60 commits.}

We plan to work on an automated continuous integration bot for behavioral change detection that will:
1) check if a behavioral change is already specified in a commit (\ie a test case that correctly detects the behavioral change is provided);
2) if not, execute behavioral change detection and test generation;
3) propose the synthesized test method to the developers to complement the commit.
Such a bot can work in concert with other continuous integration bots, such as bots for automated program repair \cite{repairnator}.

\balance
\bibliographystyle{abbrv}
\bibliography{references.bib}

\end{document}